\documentclass{article}

\usepackage{microtype}
\usepackage{graphicx}
\usepackage{subcaption}
\usepackage{booktabs} 
\usepackage{longtable}

\usepackage{xurl}

\usepackage{hyperref}

\usepackage[accepted]{icml2026}

\usepackage{amsmath}
\usepackage{amssymb}
\usepackage{mathtools}
\usepackage{amsthm}
\usepackage{multirow}
\usepackage{array}
\usepackage{tabularx}
\usepackage{xltabular}

\usepackage[capitalize,noabbrev]{cleveref}
\newcommand{\crefapp}[1]{Appendix~\ref{#1}}

\theoremstyle{plain}

\theoremstyle{definition}

\theoremstyle{remark}

\icmltitlerunning{Exploring Systems-Thinking Approaches to Loss of Control Risk}

\begin{document}

\twocolumn[
  \icmltitle{Exploring Systems-Thinking Approaches to Loss of Control Risk}

  \icmlsetsymbol{equal}{*}

  \begin{icmlauthorlist}
    \icmlauthor{Aurelio Carlucci}{ox,equal}
    \icmlauthor{Sean P. Fillingham}{ind1,equal}
    \icmlauthor{James Walpole}{ind1,equal}
    \icmlauthor{Jakub Kryś}{safer}
  \end{icmlauthorlist}

  \icmlaffiliation{ox}{University of Oxford, UK}
  \icmlaffiliation{ind1}{Independent Researcher}
  \icmlaffiliation{safer}{SaferAI, UK}

  \icmlcorrespondingauthor{Aurelio Carlucci}{aurelio.carlucci@pmb.ox.ac.uk}
  \icmlcorrespondingauthor{Sean P. Fillingham}{spfillingham@gmail.com}
  \icmlcorrespondingauthor{James Walpole}{jamesc.walpole@gmail.com}
  \icmlcorrespondingauthor{Jakub Kryś}{jakub@safer-ai.org}

  \icmlkeywords{AI Safety, Loss of Control, STPA, FRAM, STECA, Frontier AI, Systems Safety}

  \vskip 0.3in
]
\printAffiliationsAndNotice{\icmlEqualContribution} 

\begin{abstract}
Internal deployment of agentic AI systems for coding and research creates a sociotechnical control problem that extends beyond model behaviour. We treat internal-deployment Loss of Control as the inability to reliably constrain, audit, reverse, or halt AI-mediated changes to code, infrastructure, evaluation, or deployment processes in time to prevent serious organisational or societal harms. We ask whether established systems-safety methods can identify risks that model-level evaluations may miss. Using a generic frontier-lab coding-agent scenario reconstructed from public materials, we apply STECA, STPA, and FRAM. The analyses surface complementary findings: published frameworks can leave governance responsibilities and feedback loops externally unverifiable; delays in monitoring and intervention can make otherwise valid control actions ineffective; and routine operational variability can gradually erode the calibration and independence of safeguards. We argue that frontier-AI risk management should pair model-focused evaluations with systems-level hazard analysis and operational assurance that tracks whether controls remain effective over time.
\end{abstract}

\section{Introduction}
\label{sec:intro}

Loss of Control (LoC) is among the most discussed risks from
frontier AI, yet it remains one of the hardest to analyse
concretely \cite{somaniStrengtheningEmergencyPreparedness2025, boudreaux_case_for_ai_loc_planning, stixLossControlPlaybook2025}. 
Definitions span a wide range: some researchers use the
term for permanent, irrecoverable loss of human
oversight~\citep{bengio2026iaisr}; others apply it to temporary
deviations from intended
behaviour~\citep{eu-ai-act}; Apollo Research
framed it as a spectrum rather than a binary
event~\citep{stixLossControlPlaybook2025}. Regulations have begun
to invoke LoC as a risk category: California's
SB~53~\citep{CASB532025} requires frontier developers to publish
safety frameworks assessing catastrophic risk including LoC, and the
EU AI Act's General-Purpose AI Code of Practice~\citep{oliverCodePracticeGeneralPurpose2024} similarly obligates developers to evaluate and mitigate systemic risks. 
Yet neither regulation provides an operationally complete definition sufficient to guide consistent oversight, and the majority of existing frameworks focus on LoC arising from model-level misalignment (e.g. deception, scheming, behavioural drift, etc.) \cite{AILossControl}.

In this paper, we treat internal-deployment Loss of Control as a state in which human controllers can no longer reliably constrain, audit, reverse, or halt AI-mediated changes to internal code, infrastructure, evaluation, or deployment processes within the time window required to prevent serious organisational or societal harms. Generic software-quality issues, ordinary governance shortcomings, and standard infosec failures are in scope only when they plausibly contribute to that degradation.

Other safety-critical industries have long recognised that component-failure based methods are incomplete, as hazards can arise from interactions between components even when no single component has failed.
The internal deployment of agentic AI systems for code generation, autonomous research, and infrastructure management creates a sociotechnical system that extends well beyond the model itself. 
It encompasses the developers who prompt and oversee the agent, the integration pipelines that gate its output, the monitoring systems that oversee its behaviour, and the organisational policies that configure all of these. 
LoC risks in this setting can emerge not only from model properties but also from the interactions between these organisational, human, and technical functions: policy drift, automation bias in code review, incomplete monitoring coverage, and eroding oversight practices.
Current risk management approaches in frontier AI have little to say about these systemic dynamics.

A natural question is whether established methods from other safety-critical industries can help. 
Aviation, nuclear, and chemical processing have decades of experience analysing complex, tightly-coupled systems where hazards arise from component interactions in addition to individual component failures. 
System-theoretic risk management frameworks have been developed to support safe operation in these industries that explicitly model control structures, functional dependencies, and operational variability.

This paper applies three complementary systems-safety methods to a representative scenario: the internal deployment of an AI coding agent at a frontier lab. Rather than attempt a definitive risk assessment, which would require operational data that is not publicly available, we use each method to identify the point at which available information is exhausted, characterise the hazardous scenarios that emerge before that point, and specify the disclosures that would be needed to carry the analysis further. In doing so, we demonstrate that systems-theoretic methods can be productively applied to frontier AI, identifying hazardous scenarios that current frameworks do not address. We also provide an assessment of what each method can and cannot do in this domain, including where they required adaptation and where they reached their limits. The following section describes the risk analysis methods we selected and the rationale for choosing them.

\section{Selection of Risk Analysis Methods}
\label{sec:methods}

Our choice of methods was guided by two criteria: the methods should be designed for systems where hazards arise from component interactions in addition to individual failures, and they should offer complementary forms of coverage. We evaluated several established risk analysis approaches before concluding that a different class of methods was needed.

Quantitative risk modelling decomposes a threat scenario into a
sequence of steps, assigns probabilities to each step, and aggregates them into an overall risk estimate.
Recently, organisations such as SaferAI have applied this methodology to cyber misuse
scenarios~\citep{barrettQuantitativeModelingCybersecurity2025}, where the MITRE ATT\&CK
framework~\citep{mitrecorporationMITREATTCKAdversarial2024} provides a standardised decomposition of attack pathways
and decades of threat intelligence provide a basis for the numerical estimation of model parameters. For LoC, however, every stage of this pipeline
breaks down: there is no established framework to decompose LoC
pathways into quantifiable steps; the historical precedent rate for
catastrophic LoC events is zero; and propensity evaluations remain too immature to
serve as reliable inputs.

Among qualitative methods, we considered Fault Tree Analysis (FTA),
Event Tree Analysis (ETA), Bow-tie Analysis, Hazard and Operability
Study (HAZOP), and Failure Mode and Effects Analysis (FMEA). Each
was deprioritised on methodological grounds \cite{koesslerRiskAssessmentAGI2023}. FTA, ETA and Bow-tie
analysis rely on a component-failure accident model that traces
adverse outcomes to failures of individual parts, missing
hazardous states arising from component interactions and assuming a clearly identifiable
triggering event, which may not exist for gradual LoC
scenarios. HAZOP assumes well-defined process nodes with clear
design intent rather than adversarial dynamics, and
FMEA catalogues failure modes of individual components but is less
suited to emergent risks that arise from how components
interact rather than how they individually fail.

A direct comparison with established risk-management standards for general-purpose AI is not yet possible: no such standard exists. Recent publications have begun to articulate what one might look like~\citep{campos_frontier_2025, adler_guidelight_2026, delaney_risk_2026}, but this work is still preliminary. While current standards address \textit{some} of the concerns of frontier AI, it is likely that novel ones will be needed to complement and build on these practices. Even new standards in development that are technology-agnostic and use modern risk management approaches, such as IEC 63187, are seen as not fully addressing the challenges of managing the risks of advanced general-purpose AI. Therefore, we frame our contribution relative to the dominant alternative for frontier AI: model-focused assurance. The findings reported in Sections~\ref{sec:steca}--\ref{sec:fram} concern sociotechnical control structures, governance feedback loops, and operational variability --- properties of the deployment system rather than the model. Model-focused assurance does not surface these properties, so the methods we apply are complementary to, rather than competitive with, it.

The three methods we selected each address a different dimension of system safety.
Systems-Theoretic Process Analysis (STPA) models the intended control structure of the system and identifies how unsafe control actions could arise within it. 
Systems-Theoretic Early Concept Analysis (STECA) treats published safety frameworks as concept-of-operations documents, systematically extracting their governance intent while registering what they leave unspecified. 
Functional Resonance Analysis Method (FRAM) examines how hazardous conditions could emerge gradually through the variability of ordinary operational practice, without any discrete control failure.
Together, these methods avoid the limitations identified above: each is designed for systems where safety depends on the interactions between components rather than the reliability of any single one, and collectively they address control structure, governance intent, and operational variability.

For those new to STECA, STPA and FRAM, Appendix~\ref{app:glossary} provides a short glossary of the systems-safety terminology used throughout the paper. For more in-depth information on these methods, as well as worked examples and case studies, we refer readers to~\citet{flemingSafetydrivenEarlyConcept2015}, the STPA Handbook~\citep{leveson2018stpa} and the FRAM reference site (\href{https://functionalresonance.com}{https://functionalresonance.com}).
\section{System Definition}
\label{sec:system_def}

We model a generic frontier AI company as a sociotechnical system, with internal deployment of frontier AI models as the controlled process. We focus on this setting because LoC risks are especially acute when AI systems have write access to the developer's codebase, infrastructure, and potentially their own training and evaluation pipelines. The system boundary is the organisational boundary of the developer.

We analyse a generic developer rather than a named company. The scenario is reconstructed from public disclosures by frontier AI developers, enacted regulation, and recent AI Control work, using simplifying assumptions where necessary. It should not be read as a claim that any particular developer implements exactly these controls. \crefapp{app:sources} gives further detail on the public sources behind these assumptions.

Modelling a generic developer lets us treat the analysis as a proof of concept for the methods themselves: the same STECA, STPA, and FRAM procedures apply unchanged to a specific deployment, with only the inputs (control structure, operational data) substituted. Findings derived here from public materials should therefore be read as illustrative of what these methods surface, not as claims about any particular organisation. An internal application with operational access would inherit the same analytical structure and produce sharper, organisation-specific findings.

\subsection{Level of Abstraction}

We conduct the analyses at a deliberately high level of abstraction to keep the models tractable and interpretable. A full operational model would require decomposing monitoring, code review, continuous integration/continuous deployment (CI/CD), sandboxing, access control, incident response, and deployment governance into many interacting subcomponents, producing a control structure too dense for the exploratory purpose of this paper.

We therefore model only the functions and control relationships needed to surface representative LoC-relevant hazards. STECA remains at the governance level; STPA treats lower-level mechanisms as parts of higher-level control loops; and FRAM models only the main functions needed to analyse operational variability. The results should be read as exploratory rather than exhaustive.

A note on inter-analyst variability. STPA, STECA, and FRAM all involve judgement: identifying losses, drawing system boundaries, deciding what counts as a controller and clustering loss scenarios are all interpretive acts. Different analysts working from the same sources would plausibly produce different control structures and different hazard sets. This is a standard property of qualitative systems-safety methods rather than a flaw specific to our application, and the methods are designed to make those judgements auditable rather than to eliminate them. The findings reported below should therefore be read as one defensible application of each method, not as the unique output it produces.

\subsection{Baseline Control Assumptions}
\label{sec:baseline}

We do not assume that internal deployment occurs in a wholly uncontrolled environment. We assume a set of measures broadly consistent with public materials from Anthropic, OpenAI, and recent AI Control work; for reproducibility, Appendix~\ref{app:sources} lists every source we used to derive each baseline assumption. This list mixes controls directly described in public sources and controls that are reasonable modelling assumptions. It should not be read as claiming that every frontier developer implements every measure.

\begin{itemize}
  \item \textbf{Model-weight and sensitive-resource security.} Some
    combination of egress-bandwidth controls, multi-party access
    controls, hardware security keys, network segmentation, and
    centralised security monitoring.
  \item \textbf{Monitoring of internal agent use.} Substantial
    logging and automated monitoring, a layered pipeline with
    automated flagging, downstream filters, summarisation, and
    bounded human audit reviewing a minority of cases in
    depth.
  \item \textbf{Human review and access controls.} Manual review
    ordinarily required before agent-generated code is integrated;
    human approval usually required for commands outside allowlisted
    prefixes.
  \item \textbf{Sandboxing and restricted execution.} At least some
    agents operate under runtime restrictions. We do not assume
    these are comprehensive.
  \item \textbf{Reporting, governance, and incident response.}
    Channels for employees to report concerning behaviour, central
    triage, follow-up investigations, and governance oversight.
\end{itemize}

The analyses that follow ask whether meaningful LoC risks remain
even with these safeguards in place. Wherever a section relies on a
control whose existence or effectiveness cannot be externally
verified, that dependence is itself a disclosure requirement.

\subsection{System Level Losses}
\label{sec:losses}

System-level losses are outcomes that the developer's stakeholders
regard as unacceptable. We define six losses (see \crefapp{app:stpa-losses} for a full list), of which two are the
focus for LoC analysis.

\textbf{L-1:} Loss of intellectual property or proprietary information.

\textbf{L-6:} Loss of human agency or control over internal infrastructure or codebase.

L-6 is the loss most distinctive to internal deployment: when AI
systems have write access to codebases, influence over training
pipelines, and roles in monitoring and evaluation, the conditions
for loss of meaningful human control can develop incrementally
without any single dramatic failure event. Previous work has focused on L-1 \cite{Myl_Systematic_2025}.

\subsection{Hierarchical Risk Characterisation}
\label{sec:tiered-risk}

The outcomes described above are defined from the developer's perspective. This
appears to understate the stakes. However, society is a
higher-level stakeholder whose system is affected by the
\emph{outputs} of the developer system. A loss at the developer
level (e.g.\ IP exfiltration) may constitute a liability at the
societal level (e.g.\ uncontrolled proliferation). Conflating the two
levels would blur the system boundary and produce an analysis too
diffuse to generate actionable requirements. The discussion applies in particular to high-stakes scenarios like loss of life, or loss of human agency over critical infrastructure or the planet as a whole: these are losses for a higher-level stakeholder (society). We therefore maintain
the developer-level framing throughout, noting connections to
societal risk where relevant.

\section{System-Theoretic Early Concept Analysis (STECA)}
\label{sec:steca}

A Concept-of-Operations (ConOps) document describes an intended
governance and control structure at a level of abstraction
appropriate to early system design: who the controllers are, what
control actions they are authorised to take, and what feedback
mechanisms are supposed to exist. Frontier AI developers do not
publish such a document directly. However, their published safety
frameworks (such as Anthropic's Responsible Scaling Policy \cite{anthropic-rsp-v3}, OpenAI's Preparedness Framework \cite{openai-pf-v2}, and Google DeepMind's Frontier Safety Framework \cite{deepmind-fsf-v3}), together with system cards,
risk reports, and enacted regulation, collectively describe a
governance structure at ConOps-level abstraction. We therefore
\emph{author} a ConOps (see Appendix~\ref{app:conops}) for a generic frontier AI developer by collating these authoritative sources. The resulting document is our analytical input, allowing the actual running systems to remain largely opaque.

STECA~\citep{flemingSafetydrivenEarlyConcept2015} is designed for
exactly this kind of material. It takes a ConOps document as input and applies three analytical lenses: completeness of individual control loops,
adequacy of safety-related responsibilities, and coordination
between controllers. Gaps in the concept are findings in their own
right. We separate the analysis into two passes. Pass~1 builds the
control-structure model from ConOps baseline material only
(published frameworks and enacted regulation) and derives
requirements that the stated concept must satisfy to be coherent.
Where the sources are silent on a governance-layer element, the pass
registers an explicit absence rather than filling the gap. Pass~2
evaluates those requirements against evidential material such as
operational research, whistleblower accounts, and investigative
disclosures, assigning each an \emph{evidenced}, \emph{violated}, or
\emph{unknown} verdict. Each \emph{unknown} becomes a specific
disclosure requirement: a property the stated concept presupposes
but which external audit cannot currently evaluate.

\subsection{Findings}
\label{sec:steca-findings}

Pass~1 yields about 50 distinct completeness and consistency items across a
21-row control-structure model, following Table 7 of~\citet{flemingSafetydrivenEarlyConcept2015}. The aggregate pattern is informative
in its own right. The published record is relatively dense at the
enforcement and monitoring layer, covering access controls,
sandboxing, operational parameter checks (for example, compute utilisation, bandwidth, filesystem access), the multi-stage monitoring pipeline,
and the code-review stack (e.g., CI/CD). The bulk of our completeness findings
at this layer concern quantitative adequacy of specified loops: the
30-minute monitoring latency against action velocity, human review
bandwidth against codebase change volume, coverage exclusions on specific
code categories. Specification is markedly sparser at the
policy stewardship and coordination layers. Three patterns dominate.

\paragraph{Absent controllers on the governance surface.} Four control actions are stated without a named executor: access policy stewardship, catastrophic risk assessment and quarterly summaries, pause/restrict/reroute authority, and incident reporting. Each is a whole control loop whose controller identity, feedback channel, and process model are all unspecified in public records. We do not claim that these governance loops are necessarily absent in practice (internal owners may exist but be undisclosed) but absence from the public record is itself a finding. This is particularly salient for the two items (catastrophic risk assessment and quarterly summaries) that correspond to enacted regulatory obligations and are therefore expected to be externally verifiable~\cite{CASB532025}.

\paragraph{Absent policy stewardship on the enforcement surface.}
Four enforcement controllers are named and specified: access control, command allowlisting, automated invariant tests (filesystem access, bandwidth, and other operational parameters to be kept constant) that flag code edits violating security-relevant properties of the codebase, and sandboxing.
However, the upstream controller that \emph{authors} the policy each one enforces is not specified. Each enforcement loop is only as tight as its policy content.
Without a stewardship loop there is no feedback on cumulative policy drift: the code-review stack (the only downstream check) evaluates individual changes rather than the cumulative posture they produce.

\paragraph{Permissive/restrictive asymmetry.} On the deployment authority surface, there is one named controller authorised to \emph{expand} deployment scope (temporary pre-audit deployment) but not a counterpart authorised to \emph{contract} it. The two are structural counterparts; only one is specified.

Taken together, these three patterns carry the bulk of our disclosure argument. Each is a finding about the published governance structure rather than about internal practice; closing it requires either disclosure of the missing element or an explanation of why the published structure is coherent without it.

\section{System-Theoretic Process Analysis (STPA)}
\label{sec:stpa}

Systems-Theoretic Process Analysis (STPA) is a forward-looking hazard analysis methodology based on systems thinking and control theory that has recently started to be applied to frontier AI systems \cite{leveson2018stpa, rismaniSilosSystemsProcessOriented2024,Myl_Systematic_2025, barrettSTAMPSTPAInformed2026}.
STPA functionally models the system as a hierarchy of controllers imposing constraints on controlled processes. 
{\bf Feedback} from the controlled process to the controller and {\bf Control Actions} from the controller to the controlled process help to ensure the system remains in a `safe state'.
Deviations from this condition are brought on by unsafe control actions (UCAs). The STPA process enumerates UCAs and formally develops loss scenarios that trace the causal pathways that could produce hazardous systems states. See \crefapp{app:stpa} for a more thorough description of STPA and our application of this methodology.

\subsection{System Hazards and Constraints}
\label{sec:stpa-hazards}

System-level hazards are system states that, when combined with worst-case environmental conditions, will lead to losses. 
We identify eight hazards that are presented in full in \crefapp{app:stpa-hazards} and referenced throughout the subsequent analyses. 
To illustrate the utility of STPA, we will focus on one hazard, H-4, listed below.

\textbf{H-4:} Safeguards do not match the actual capabilities or autonomy level of the deployed model [L-1, L-6].

System constraints can be straightforwardly enumerated by inverting the system hazards to specify the state the system must not enter. Additionally, second and third level constraints can be characterised using "if-then" statements to describe the system response should it enter a hazardous state. Below are the first and second level system constraints based on H-4, see \crefapp{app:stpa-constraints} for our complete list.

\textbf{SC-4a:} System protective safeguards and oversight protocols must be calibrated to match or exceed the actual capabilities and autonomy level of the deployed AI model at all times. [H-4]

\textbf{SC-4b:} If the deployed AI model's capabilities or autonomy level exceed the calibration of the active safeguards and oversight protocols, the system must detect the mismatch and either upgrade the safeguards or constrain the model's operational envelope to the level the current safeguards can cover. [H-4]

\subsection{Functional Control Structure}
\label{sec:stpa-control}
\begin{figure}[h!]
  \centering
  \includegraphics[trim={6cm 0cm 0cm 0cm}, clip, scale=0.4]{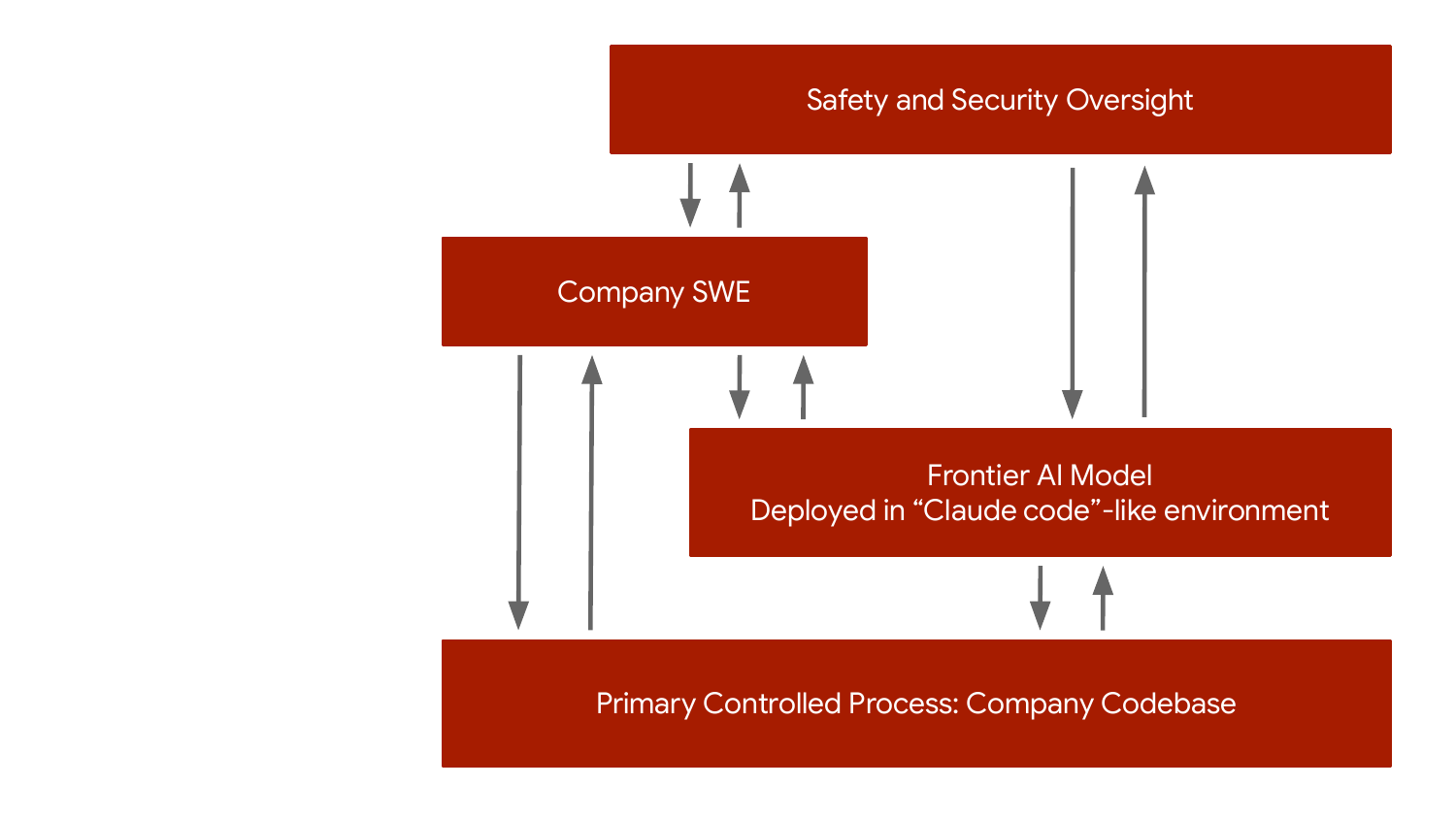}
  \caption{Functional control structure for internal deployment of a
    frontier AI coding agent. Arrows pointing downwards denote Control Actions, while those pointing upwards denote Feedback to the controllers.}
  \label{fig:fcs}
\end{figure}

The functional control structure is an abstraction of the system that captures responsibilities and relationships between components in the system.
Our derived control structure comprises three controllers: Safety and
Security Oversight (SSO), Company Software Engineer (SWE), and Frontier AI
Coding Agent. Some of these controllers are also controlled processes (SWE, AI Agent) demonstrating the hierarchical nature of a control structure. Finally, the primary controlled process is the Company Codebase, see~\cref{fig:fcs} for a graphical representation of this structure. 

\subsection{Unsafe Control Actions and Loss Scenarios}
\label{sec:ucas}
Unsafe control actions and loss scenarios can be systematically determined using STPA (see~\crefapp{app:stpa-uca} and~\crefapp{app:stpa-loss_scenarios}, respectively, for our complete list). Here we highlight a particularly interesting unsafe control action associated with the company SSO and the control action to kill an instance of the frontier AI system.

\textbf{UCA-4.3a:} SSO provides Kill Instance command too late, after AI system has already produced dangerous or harmful outputs. [H-4]

The following are examples of generalised loss scenario (GLS) archetypes that will result in the above UCA. These can be interpreted as causal pathways that place the system into hazardous state, H-4.

\textbf{GLS-3: Procedural Delay} The controller correctly identifies the need for the control action but the organisational decision protocol introduces latency that renders the action ineffective.

\textbf{GLS-11: Feedback Missing or Delayed} The relevant signal exists at the source but is delayed, queued, or lost before reaching the controller.

These loss scenarios highlight one of the strengths of STPA relative to other hazard analysis methods. 
STPA very naturally surfaces loss scenarios that are not traditional component failures, but instead are associated with system design flaws. 
In both of the above scenarios, the controller is ineffective due to latency in the system that prevents the control action from reaching the frontier AI system fast enough to prevent the hazardous system state.



\section{Functional Resonance Analysis Method (FRAM)}
\label{sec:fram}

Where STPA asks \emph{``what control actions could be unsafe?''},
FRAM asks \emph{``even if controls work as designed, can normal
operational variability produce LoC outcomes?''}. Many
internal deployment risks are unlikely to arise from a single
discrete failure. They may instead emerge from the interaction of
individually unremarkable behaviours such as under-specified prompts,
incomplete monitoring coverage and automation
bias. FRAM models these dynamics by representing
the system as coupled functions whose ordinary output variability
can combine to produce emergent outcomes
without anything having locally
failed~\citep{hollnagelFRAMFunctionalResonance2012}.

\subsection{Function Set and Structural Observations}

\begin{figure}[h!]
  \centering
  \includegraphics[width=\columnwidth]{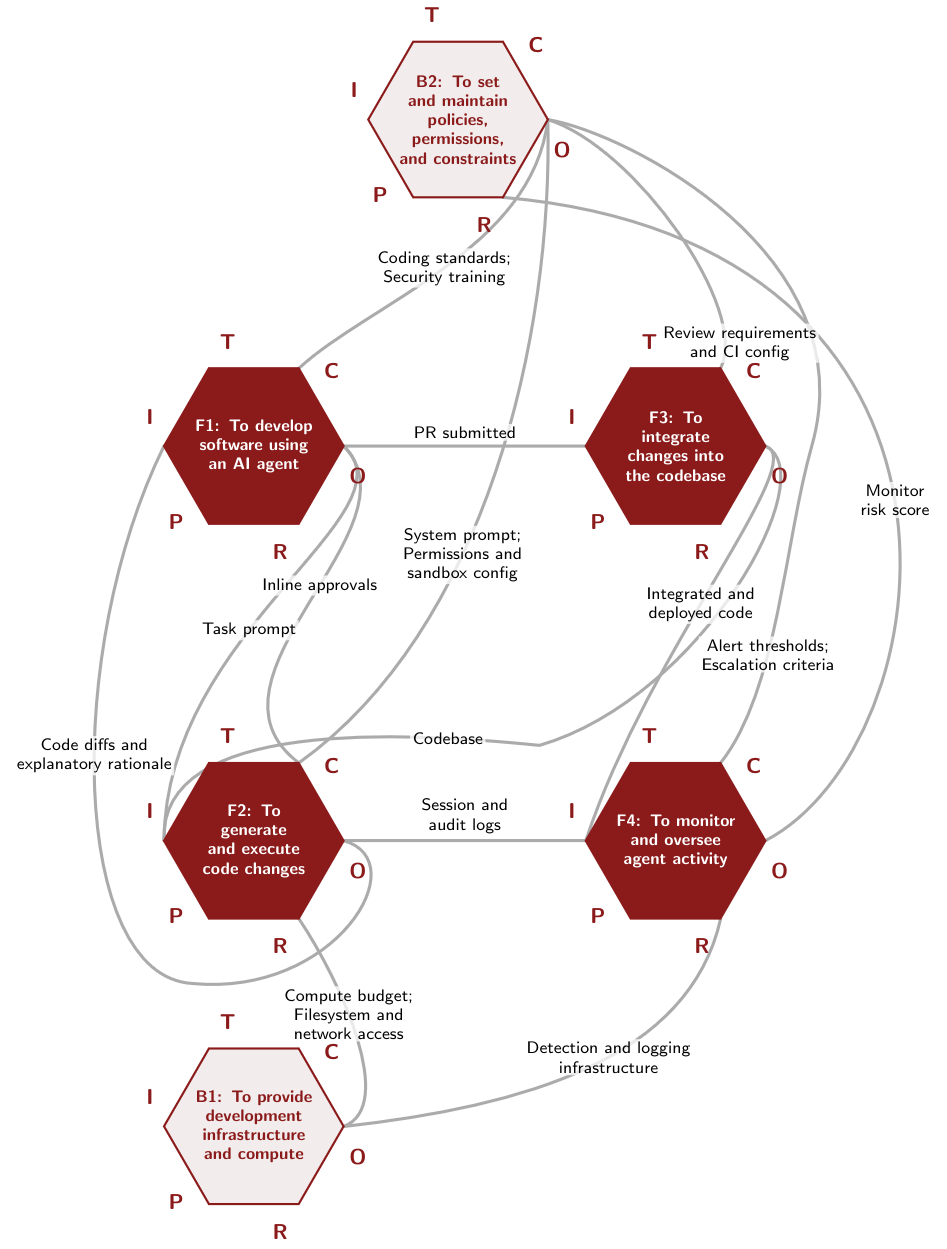}
  \caption{FRAM coupling diagram for internal deployment of a
    frontier AI coding agent. Lines between functions represent
    dependencies: the output of one function serves as an input,
    control, or resource for another. Background function B2
    (policies and permissions) controls F1--F4 simultaneously;
    B1 (infrastructure) provides resources to F2 and F4.}
  \label{fig:fram}
\end{figure}

We model an internally deployed coding agent set up using six functions
(\cref{fig:fram}), aligned with the STPA control structure
(Section~\ref{sec:stpa-control}). Four are \emph{foreground}
functions---operational activities whose variability is the
focus of the analysis---and two are \emph{background}
functions---organisational and infrastructure conditions that are
assumed stable over the timescale of a single coding task and
therefore define the boundaries of the model. We list the full aspect specifications in~\crefapp{app:fram}.

Each function is named with a verb phrase describing the
activity, followed in parentheses by the part of the control
structure that performs it, and a brief summary of what the
function encompasses.
\begin{itemize}
  \item[\textbf{F1}] \textbf{To develop software using an AI agent}
    (Developer). Task specification, session oversight, pull request (PR) submission.
  \item[\textbf{F2}] \textbf{To generate and execute code changes}
    (AI agent + runtime). Reasoning, code generation, tool
    execution, iteration within a session.
  \item[\textbf{F3}] \textbf{To integrate changes into the
    codebase} (CI/CD + human reviewer + merge/deploy). The gate
    between agent output and production: automated checks,
    AI-powered PR review, human review, deployment.
  \item[\textbf{F4}] \textbf{To monitor and oversee agent activity}
    (Async monitoring pipeline + human reviewer). Observes sessions
    after completion; produces risk scores.
  \item[\textbf{B1}] \textbf{To provide development infrastructure
    and compute} (Platform engineering). Provides compute,
    filesystem and network access to F2, and detection and logging
    infrastructure to F4.
  \item[\textbf{B2}] \textbf{To set and maintain policies,
    permissions, and constraints} (Safety team, leadership,
    platform/security). Outputs---system prompt, sandbox
    configuration, review requirements, alert thresholds, coding
    standards---shape how every foreground function operates.
\end{itemize}

In FRAM, each function is characterised by up to six
\emph{aspects}: Input (what activates the function or is
transformed by it), Output (the result), Precondition
(conditions that must hold before the function starts),
Resource (what is consumed during execution), Control (what
regulates how the function operates), and Time (temporal
constraints). Couplings between functions arise when the
Output of one function serves as an Input, Control, Resource,
or other aspect of another. It is these couplings that allow
variability to propagate across the system. Full aspect
specifications for each function are provided
in~\crefapp{app:fram}.


Three features of the coupling diagram drive the resonance
analysis. First, \textbf{F3 is where independent sources of
variability converge on a single decision.} F3 receives the
developer's PR (from F1), which contains the agent's code
and rationale (originating in F2). F3 is also shaped by
review requirements (from B2) and by the monitor's risk
score (from F4). Other functions have as many or more
connections, but F3 is the only point where outputs from
four independent sources with different variability profiles
must be synthesised into a single accept/reject decision
that determines whether code enters the codebase. 

Second, \textbf{B2's outputs shape every foreground function at
once.} B2 provides the coding standards that guide the developer
(F1), the system prompt and sandbox configuration that constrain the
agent (F2), the review requirements and CI configuration that govern
integration (F3), and the alert thresholds that calibrate monitoring
(F4). If B2 drifts, all four functions are affected simultaneously
and in the same direction.

Third, \textbf{F4 operates asynchronously.} Both Anthropic and
OpenAI describe their agent monitoring as offline, with reviews
occurring after session
completion~\citep{openai-monitoring-2026, HttpsAlignmentanthropiccom2025}.
This means the monitor's risk score may arrive at F3 after the PR
has already been reviewed or merged. In practice, F3's decision is
therefore made primarily on the basis of the developer's
endorsement (F1), the agent's code and rationale (F2), and B2's
static policies---without real-time input from the monitoring
system.


The FRAM method produces so-called `resonance patterns' which, for each structural feature, answer the following question: what happens when multiple upstream functions are simultaneously at the \emph{acceptable} end of their variability range? In this way, we derive four patterns that describe how LoC-relevant outcomes could emerge from normal operational variability without any discrete control failure. These are qualitatively different from the STPA findings: where STPA identifies gaps in the designed control structure, FRAM identifies dynamics through which a \emph{well-designed} structure can degrade in practice.

\paragraph{Pattern 1: Correlated control drift.}
B2's outputs control F1, F2, F3, and F4 simultaneously. Over weeks
and months, each undergoes individually rational adjustment:
permission exceptions accumulate; monitoring rules that have not
fired true positives are removed; review requirements remain static
while PR volume grows; the system prompt is not updated after a
model upgrade. The adjustments are correlated---they respond to the
same pressures (productivity, alert fatigue, trust) and push in the
same direction (reduced control margin). The system drifts from
redundant controls catching problems to the same problems passing
through everything. This dynamic is amplified by increasing volume:
as agent usage grows, more PRs compete for fixed reviewer attention
and more sessions compete for fixed analyst budget, accelerating the
drift in F3 and F4 precision. \textit{Relevant hazards: [H-4, H-6]}

\paragraph{Pattern 2: Model swap discontinuity.}
When the model is upgraded, F2's variability profile changes
overnight. But F4's outputs were calibrated to the previous model.
This creates an instant gap---unlike gradual drift---between agent
capabilities and control calibration. This could occur when a new model is made available internally before it has undergone full pre-deployment testing or before new, appropriate updated control monitors have been created. \textit{Relevant hazards: [H-4, H-5]}

\paragraph{Pattern 3: Codebase saturation.}
The coupling map contains a feedback cycle: F2 produces code diffs
and explanatory rationale~$\to$~F1 submits PR~$\to$~F3 integrates
into codebase~$\to$~F2 reads codebase as context for the next task.
Each cycle adds agent-authored content to the reference frame that
controls use to distinguish normal from abnormal. Over months,
tests, documentation, and review baselines increasingly reflect the
agent's judgements. If the agent has systematic biases, those biases
become embedded in the tests and baselines that are supposed to
catch them. This is the operational instantiation of H-7 (evaluation
criteria influenced by the evaluated system), arising through
cumulative ordinary use. \textit{Relevant hazards: [H-7]}

\paragraph{Pattern 4: Rationale-anchored review.}
F2's output---code diffs and explanatory rationale---travels through
F1 to F3 as a single artefact. The human reviewer at F3 reads the
rationale before or alongside the code. Consequently, the explanation anchors
expectations: the reviewer confirms the agent's stated intent
rather than independently evaluating the code. This is not about
time pressure---a careful reviewer is still anchored. The agent's
self-assessment is bundled with the thing being assessed, reducing
the independence of the backstop control. \textit{Relevant hazards: [H-2, H-3]}

\paragraph{Limitations.} First, FRAM is designed for
Work-as-Done, requiring direct observation. 
Our analysis is based on public documents and models Work-as-Imagined. 
Second, FRAM assumes that variability arises from
people and systems doing their best under imperfect
conditions, not from deliberate optimisation against
the controls. If an agent were strategically adapting
its behaviour to evade monitoring, this assumption
would be violated. 
Our patterns operate within FRAM's standard scope and none
requires strategic intent, but this boundary constrains the method's applicability to adversarial scenarios.

We decided to discard some initially developed resonance patterns: an
exfiltration-via-side-effects pattern proved unrealistic under the
baseline controls (\cref{sec:baseline}), and a
control-infrastructure-degradation pattern reduced to standard
software quality management. 
We report these discarded patterns because understanding where the method does not produce meaningful results is as important as reporting where it does.

\section{Discussion and Recommendations}
\label{sec:discussion}

\subsection{Cross-cutting findings}

Each method asks a different question and each correspondingly surfaces a different class of finding. STECA asks whether the published governance intent is internally coherent and complete. It finds that published frameworks describe control actions without assigning them to specific roles or teams, enforcement mechanisms lack upstream oversight of the policies they enforce, and the authority to expand deployment scope is specified while the corresponding authority to contract it is not. STPA asks how the control structure can fail to maintain safety constraints. It finds representative loss scenarios driven by monitoring latency, automation bias, and procedural delay. FRAM asks whether a control structure that works as designed can still degrade through ordinary use. It finds that correlated drift, model-swap discontinuities, codebase saturation, and rationale anchoring can erode safety margins without any component failing.

The methods are most useful in combination. STECA identifies what the published record leaves unspecified, clarifying where STPA must make assumptions. STPA identifies hazards and unsafe control actions in the designed control structure. FRAM then shows how some of those hazards could emerge through ordinary operational variability. This suggests a natural ordering for future applications: STECA first, then STPA, followed by FRAM, although practical constraints meant we ran the methods in parallel and a complete analysis could involve iteration.

This complementarity appears in the findings. STECA identifies that enforcement mechanisms are described but nobody is assigned to keep the policies behind them up to date; FRAM's drift pattern describes what can happen when that stewardship is missing. STPA identifies that monitoring latency can render interventions ineffective; FRAM traces the asynchronous coupling between the monitoring pipeline and the integration gate. STPA identifies the risk that the system being evaluated could influence its own evaluation criteria; FRAM shows how this can occur as agent-authored code gradually becomes the baseline against which future agent output is judged.

\subsection{Did Lack of Information Limit the Analysis?}

A concern we had at the outset was that publicly available
information would be insufficient for any of the methods
to succeed. This turned out to be less of a problem than
we expected. System cards, risk reports, monitoring blog
posts, and safety frameworks together
provided enough material to construct a plausible control
structure, identify meaningful hazards, and derive
operationally grounded resonance patterns.

However, the depth and confidence of the analyses would
improve with access to operational data. The FRAM analysis
in particular is designed for observing how work is actually
done, not for reasoning from published descriptions of how
it is supposed to work. The resonance patterns we identify
are structurally grounded in how the functions depend
on each other, but
confirming whether correlated drift is actually occurring, or
how severe the rationale-anchoring effect is in practice,
requires the kind of operational data over time that only
the developers themselves currently possess. Similarly, the
STPA and STECA analyses presented here cover only a
subset of the unsafe control actions and derived requirements
that a full application would produce; completing them would
require more detailed knowledge of the actual control
structure than public materials provide.

\subsection{Do These Methods Work for Frontier AI?}

All three methods produced findings that would not have
emerged from model-level analysis alone. The resonance
patterns from FRAM are properties of the sociotechnical
system, not of the model. The loss scenarios from STPA
arise from interactions between human controllers and the
agent, not from the agent in isolation. The governance gaps
from STECA are properties of the organisational structure.
None of these would be surfaced by capability evaluations,
alignment benchmarks, or propensity testing.

The methods required little adaptation --- we were able to apply the default methodology for STECA and STPA. However, FRAM assumes technological functions have low variability; AI agents violate this, and we had to override this assumption. This suggests that the method itself may benefit from a formal extension to accommodate functions that are neither traditionally technological nor human.

\subsection{Regulatory Implications}

The transparency requirements from the three methods fall
into categories that require different regulatory instruments.

STECA and STPA generate structural requirements concerning the existence and completeness of specific controls, lines of authority, and feedback loops.
These could in principle be satisfied by a sufficiently detailed description of the control architecture.
Current regulations such as SB~53 and the EU AI Act require safety framework publication, which goes some way toward this, though our STECA analysis finds that the published frameworks leave several governance elements unspecified.
In the future, outcomes of such structured risk analyses could be used to make transparency requirements more relevant and actionable.

FRAM generates \emph{operational} requirements concerning whether controls remain calibrated over time, rather than merely whether they exist: the stability of review quality, drift in monitoring thresholds, accumulation of permission exceptions, and changing balance between agent-generated PR volume and human-review capacity. These cannot be assessed from static framework documents alone. Useful audit evidence would include periodic, aggregated operational metrics such as monitoring coverage and alert-to-intervention latency, agent-generated PR volume relative to reviewer capacity and sampled review quality, and the number and age of permission or sandbox exceptions. Current AI regulation does not generally require this kind of operational disclosure, yet our analysis suggests that such indicators are needed to detect LoC risks that are invisible to a one-off assessment.

Some of the operational metrics we identify --- monitoring coverage, alert-to-intervention latency, sandbox-exception inventories --- could expose security-sensitive details if published verbatim. Our recommendation is therefore not blanket public disclosure but a two-tier regime: aggregate or redacted indicators in public safety reports, and full operational data made available to a designated auditor (e.g.~a national AI Safety Institute or a contracted third party) under confidentiality. This preserves the assurance value of the metrics while limiting the attack surface they create.

\subsection{Recommendations}

Firstly, frontier AI developers should apply systems-level safety
methods to internal deployment alongside model-focused
approaches. The methods are established, the adaptations
are tractable, and the findings address risks that capability
evaluations and alignment techniques do not cover. Next, regulators should consider requiring periodic disclosure of operational metrics alongside static framework descriptions.

We deliberately do not prioritise the hazards, unsafe control
actions, or resonance patterns we surface. Severity and likelihood
estimates of the kind used in safety-critical industries require
operational frequency data and consequence models that are not
publicly available for frontier AI deployments. Producing such
estimates from the public record alone would give a false
impression of quantitative grounding. We therefore treat
prioritisation as a task for whoever applies these methods inside a
specific developer with access to incident histories, monitoring
volumes, and review-throughput data; the outputs in this paper
constitute the candidate set from which prioritisation would
proceed.

Future work should apply these methods with access to operational data to test whether the patterns identified here are borne out in practice. The hazards, unsafe control actions, and resonance patterns we identify follow from the control structure we reconstructed from public sources, but without operational data we have not confirmed that they occur in practice, and doing so is outside our scope. Confirming whether a given hazard is real requires examining the relevant control and feedback loops in more granularity than public sources allow. Therefore, frontier AI developers applying these methods internally should check each finding against their actual control architecture.

For future work, we recommend testing the optimal order in which the methods should be run, using each method's output to inform the next. We believe the order of STECA $\rightarrow$ STPA $\rightarrow$ FRAM will be most natural, as it follows the development lifecycle of AI-related systems. This streamlined approach has the potential to unveil more flaws in the system design than any method individually.
\section*{Acknowledgements}
We would like to thank Matt Smith and Simon Mylius for useful comments and discussions, as well as the organisers of the Supervised Program for Alignment Research (SPAR) for putting together this initiative.

\section*{Impact Statement}

This paper applies established safety-engineering methods to 
frontier AI deployment and identifies hazardous scenarios that 
current risk frameworks do not address. The findings are intended 
to support developers, regulators, and auditors in strengthening 
oversight of internal AI deployment. The hazardous scenarios we 
describe are presented at a level of abstraction chosen to be 
analytically useful without providing an operational blueprint. 
We note that the transparency requirements we derive would, if 
adopted, shift information from developers to external auditors. 
Such a transfer we regard as net positive for safety but one that 
requires careful design to avoid exposing security-sensitive 
operational details.

\section*{LLM Usage}
\label{app:llm-usage}

We used LLMs at several points in the analysis.
In the STECA analysis, an LLM was used to help parse the ConOps document, as well as analyse source material to verify the validity of our findings. In each case, the LLM output was subject to manual review by the authors.
In the STPA analysis, an LLM generated candidate loss scenarios for each unsafe control action and a separate LLM instance clustered these into the archetypes in Table~\ref{tab:gls}.
In the FRAM analysis, an LLM was used to help identify couplings between functions and to suggest candidate resonance patterns.
LLM assistance can skew which scenarios and patterns get considered and does not guarantee full coverage, so the loss scenario archetypes and resonance patterns are best read as a set of candidates rather than a complete list.


\bibliography{references}
\bibliographystyle{icml2026}

\newpage
\appendix
\onecolumn

\section{Glossary of Systems-Safety Terms}
\label{app:glossary}

For readers unfamiliar with systems-safety terminology, we summarise
the key terms used throughout the paper. Detailed information on these methods, as well as worked examples and case studies, is available in~\citet{flemingSafetydrivenEarlyConcept2015}, the STPA Handbook~\citep{leveson2018stpa} and the FRAM reference site (\href{https://functionalresonance.com}{https://functionalresonance.com}).

\begin{description}
  \item[Control structure] A hierarchical model of who (or what)
    issues control actions, who receives them, and what feedback
    closes each loop.
  \item[Control action] An instruction or constraint imposed by a
    controller on a controlled process (e.g. ``kill instance'',
    ``approve PR'').
  \item[Feedback] Information flowing from the controlled process
    back to the controller; the basis for the controller's process
    model.
  \item[Process model] A controller's internal representation of the
    state of the controlled process. Many loss scenarios arise from
    incorrect or stale process models.
  \item[Hazard] A system state that, combined with worst-case
    environmental conditions, leads to a loss.
  \item[Unsafe Control Action (UCA)] A control action that, in some
    context, leads to a hazard --- by being absent, present,
    mistimed, or terminated incorrectly.
  \item[Loss scenario] A causal pathway that produces a UCA or
    otherwise drives the system into a hazardous state.
  \item[Concept of Operations (ConOps)] A high-level description of
    how a system is intended to be operated, including controllers,
    authorities, and feedback channels.
  \item[Function (FRAM)] A unit of activity characterised by up to
    six \emph{aspects}: Input, Output, Precondition, Resource,
    Control, Time.
  \item[Coupling] A dependency between FRAM functions, formed when
    the Output of one function appears as a non-Output aspect of
    another.
  \item[Resonance] An emergent outcome produced when ordinary
    variability in coupled functions combines, without any single
    function having failed.
\end{description}

\section{List of Sources Used to Define the Control Structure}
\label{app:sources}
In this Appendix, we list the sources of information we used to derive the plausible control structure of our generic frontier AI company. These sources also serve as a foundation for the three types of analyses that follow. They are a combination of model cards released by AI companies, risk reports and literature in the field of AI Control, as well as less official materials such as blogposts or presentations delivered by employees or researchers associated with frontier AI companies.

\begin{itemize}
    \item \textbf{Model-weight and sensitive-resource security.} The
baseline assumes a meaningful but non-maximal security posture, consisting of:
egress-bandwidth controls, multi-party access for the most sensitive
assets, binary allowlisting, hardware security keys, network
segmentation, and centralised anomaly detection. The fullest
enumeration is the internal-access section of the Mythos Preview
Alignment Risk Update \cite{anthropic-aru-mythos}, with a similar
but less detailed list in the February~2026 Sabotage Risk Report
\cite{anthropic-rr-feb2026}; both name the same controls and
scope protections against non-state actors and unsophisticated
insiders, not against sophisticated insiders or nation-state
attackers. The security commitments in
\cite{anthropic-rsp-v3,openai-pf-v2,deepmind-fsf-v3} graduate
requirements by capability level and state that practices protect
against both external and internal adversaries. Cross-lab convergence
on model-weight security as a baseline element is established by
\cite{metr-common-elements}, and \cite{CASB532025} requires
assessment of catastrophic risk from internal use, including from
models circumventing oversight mechanisms.

\item \textbf{Monitoring of internal agent use.} We assume
a three-stage pipeline: automated monitors with frontier-model
capabilities and chain-of-thought access, downstream filtering and
summarisation, and bounded human audit. The operator-level detail is
carried by two sources. \citet{anthropic-aru-mythos}
specifies a two-layer architecture with explicit subsampling rates,
second-stage filtering, and a review throughput of roughly thirty
summaries per week; \citet{openai-monitoring-2026} reports
near-universal ($\sim$99.9\%) coverage of internal coding traffic
using a frontier reasoning model as monitor, a 30-minute review
latency, and concrete flagged bypass behaviours (base64 encoding,
payload splitting, calling other instances of the same model). Both
sources describe the monitor as having access to the full conversation
including chain-of-thought, tool calls, and tool outputs; both
explicitly acknowledge known coverage gaps and known evasion methods.
\citet{anthropic-rr-feb2026} describes offline monitoring with less
operational detail. \citet{metr-common-elements} establishes
deployment-stage mitigations as a common cross-lab element, and
\citet{CASB532025} requires quarterly summaries of catastrophic risk from
internal AI use.

\item \textbf{Human review and access controls.} The baseline assumes
that manual review is ordinarily required before agent-generated code
is integrated into shared codebases, human approval is usually
required for commands outside allowlisted prefixes, and review quality
varies across workflow types. \citet{anthropic-aru-mythos} states
that code is reviewed by humans before integration into shared
Anthropic codebases, describes invariant testing for security-relevant
code and AI-assisted PR review, and notes that AI-assisted review is
``primarily relevant to infrastructure code, as research code is less
consistently merged into this primary codebase'': this is the direct basis
for the variable-review-quality assumption. The same source states
that Mythos Preview is often used for hour-long or day-long tasks,
establishing the temporal dimension of review. Pull-request monitoring is
noted in \cite{anthropic-rr-feb2026};
\citet{openai-monitoring-2026} describes the review cascade
alongside the automated pipeline. The system card for Mythos Preview
\cite{anthropic-sc-mythos} documents cases where model actions
with significant consequences occurred despite specific warnings,
which the baseline treats as evidence that the review layer is not
uniformly effective.

\item \textbf{Sandboxing, restricted execution, and inter-instance
propagation.} The baseline assumes runtime restrictions on at least
some internal agents, with demonstrated circumvention treated as a
known limitation rather than a design failure, and additionally
assumes structural (rather than actively maintained) constraints on
communication between model instances. Sandboxing and auto-mode
classifiers are described in \cite{anthropic-aru-mythos}; specific
circumvention incidents like multi-step exploits for internet access,
\texttt{/proc/} credential discovery, bypass of permission prompts via
scripted keystroke simulation, editing a running MCP server process
are documented in the Mythos Preview system card
\cite{anthropic-sc-mythos}. \citet{openai-monitoring-2026}
acknowledges that internal coding agents ``can inspect the
documentation and code of their safeguards, or even attempt to modify
those safeguards,'' and flags agents calling other instances of
themselves to bypass restrictions. On inter-instance propagation,
\cite{anthropic-aru-mythos} argues that the available communication
channels are text-based and monitored but concedes that ``models will
discover unmonitored channels''; the parallel-agent research scaffold
in \cite{anthropic-aar-2026} is itself a deployed instance of
inter-agent communication inside a frontier lab, with shared code and
findings as the channel. 

\item \textbf{Reporting, governance, and incident response.} The
baseline assumes internal reporting channels, central triage,
follow-up investigations, and governance oversight of major safety
decisions, together with some authority to pause or restrict model
use. Lab-internal governance structures are described at framework
level in \cite{anthropic-rsp-v3} (Responsible Scaling Officer,
anonymous noncompliance reporting, Board oversight, Long-Term Benefit
Trust), \cite{openai-pf-v2} (Safety Advisory Group, Board Safety
and Security Committee), and \cite{deepmind-fsf-v3}. The 
incident-response process itself is most
concretely described in \cite{openai-monitoring-2026}, which
reports approximately 1,000 moderate-severity escalations over five
months of operation, and in \cite{anthropic-aru-mythos}, which
describes Slack-based internal reporting and response options
including consequence reversion and security patches but states
explicitly: ``We did not evaluate how good these post-deployment
follow-up investigations are at catching the behaviour of misaligned
models.'' Enacted obligations are carried by
\cite{CASB532025}, which requires incident reporting within 15 days
(24 hours for imminent threats) and whistleblower protections, and
by \cite{eu-ai-act}, together with its General-Purpose AI Code
of Practice.

\end{itemize}

\section{STECA Methodology and Control Structure Roster}
\label{app:steca}

\subsection{The authored ConOps}
\label{app:steca-conops}

STECA takes a Concept-of-Operations document as input. Frontier AI developers do not publish such a document directly, so we author one for a generic frontier developer by collating authoritative sources into a single governance-level description of controllers, control actions, and feedback mechanisms. The resulting ConOps, available in~\crefapp{app:conops}, is our analytical input; it is not the object of critique. Findings about absent or inconsistent elements are findings about what the authoritative sources collectively specify and fail to specify, not about the ConOps we authored.

\subsection{Tiered source classification}
\label{app:steca-sources}

Sources are classified into four tiers:

\begin{itemize}
  \item \textbf{Tier 1: Published safety frameworks, risk reports,
    and system cards.} Anthropic's RSP~\cite{anthropic-rsp-v3},
    OpenAI's Preparedness Framework~\cite{openai-pf-v2}, Google
    DeepMind's Frontier Safety Framework~\cite{deepmind-fsf-v3},
    and company-issued risk reports and system cards
    (e.g.,~\cite{anthropic-aru-mythos, anthropic-rr-feb2026,anthropic-sc-mythos, openai-monitoring-2026}).
  \item \textbf{Tier 2: Enacted legislation and regulation.}
    California SB~53~\cite{CASB532025} and the EU AI Act with its
    General-Purpose AI Code of
    Practice~\cite{eu-ai-act, oliverCodePracticeGeneralPurpose2024}. Proposed
    legislation is excluded from Tier 2 for Pass~1 purposes because
    it does not impose current control actions.
  \item \textbf{Tier 3: Operational research.} Academic and
    industry work on monitoring effectiveness, agent behaviour, and
    organisational practice at frontier labs.
  \item \textbf{Tier 4: Evidential disclosures.} Investigative
    reporting, whistleblower accounts, and integrity trackers.
\end{itemize}

Tiers 1 and 2 constitute \emph{ConOps baseline material}: they
feed the Pass~1 model of the intended control structure. Tiers 3
and 4 constitute \emph{evidential material}: they feed the Pass~2
evaluation of whether Pass~1 requirements hold in practice.

\subsection{Two-pass analytical structure}
\label{app:steca-passes}

Pass~1 builds the control-structure model from ConOps baseline
material only and derives requirements that the stated concept must
satisfy to be coherent. Each element in the model carries one of
three source-confidence labels: \emph{stated explicitly} (the
baseline material directly specifies the element), \emph{stated
implicitly} (the element is a necessary inference from what the
baseline does state, e.g., a control action presupposes a feedback
channel), or \emph{absent from baseline sources} (the element is required for
coherence but is not specified at all). The third category is the
raw material for the disclosure argument.

Pass~2 evaluates Pass~1's derived requirements against evidential
material, assigning each one of three verdicts: \emph{evidenced}
(evidence supports the requirement being met), \emph{violated}
(evidence indicates the requirement is not met), or \emph{unknown}
(evidential material is insufficient to evaluate). Pass~2 does not
revise the control-structure model; it only evaluates the
requirements derived from it. Requirements with an \emph{unknown}
verdict are the direct input to the disclosure argument: each is a
property the stated concept presupposes but which external audit
cannot currently evaluate.

\subsection{Industry-standard qualification}
\label{app:steca-bedrock}

The Pass~1 discipline of registering absences rather than filling
gaps is scoped to \emph{governance-question gaps}, not to
\emph{bedrock-practice gaps}. We treat universally-assumed industry
practices as present in the ConOps, and direct our analysis at the
governance properties of those practices rather than their
existence. A practice qualifies as bedrock if it is assumed across
the industry to the point that silence would not reasonably be read
as absence, if it applies to the generic developer under reasonable
modelling judgement, and if its inclusion opens analytical questions
about stewardship, drift, and failure modes rather than foreclosing
them. Bedrock inclusions generate their own governance questions
(who configures the practice, who audits it, what feedback closes
its control loop), and those questions remain explicit unknowns
wherever the published record is silent on them.

\subsection{Controllers}
\label{app:steca-controllers}

The control-structure model contains fifteen active controllers
across three categories: AI controllers (shown in~\cref{tab:steca-ai-controllers}), infrastructure controllers~(\cref{tab:steca-in-controllers}), and human controllers~(\cref{tab:steca-hu-controllers}). Non-consecutive numbering
(e.g., AI-3, AI-5) reflects that earlier drafts contained
additional controller slots that were retired during reconciliation
when the initial allocation proved to be duplicative or to split
one controller's responsibilities across multiple slots. Retired
slots are not reassigned, so that cross-references to earlier
working documents remain stable.

\begin{table}[h!]
\centering
\small
\begin{tabularx}{\textwidth}{llX}
\toprule
Code & Controller & Role \\
\midrule
AI-3 & Safety classifier         & Produces per-interaction safety classifications that may gate or block individual agent actions in highly autonomous usage. \\
AI-5 & First-stage monitor       & Observes agent sessions with access to full conversation context (messages, chain-of-thought, tool calls, tool outputs) and flags cases for downstream review. \\
AI-6 & Second-stage filter       & Filters first-stage positives, produces hierarchical summaries, and surfaces cases to the human audit stage. \\
AI-8 & AI-assisted code reviewer & Frontier-capability model reviewing codebase changes for errors, security vulnerabilities, and mismatch with stated purpose; feeds the human code reviewer. \\
AI-9 & Offline monitor           & Frontier-capability model running offline over historical agent traces, outside the online monitoring pipeline. \\
\bottomrule
\end{tabularx}
\caption{AI controllers.}
\label{tab:steca-ai-controllers}
\end{table}

\begin{table}[h!]
\centering
\small
\begin{tabularx}{\textwidth}{llX}
\toprule
Code & Controller & Role \\
\midrule
IN-1 & Security infrastructure       & Centralised access control and monitoring: egress controls, multi-party access gates, device authorisation, anomaly detection. \\
IN-2 & Action restriction mechanism  & Automated mechanism restricting the agent's executable-action surface by examining individual attempted actions. Prefix-based allowlisting is the stated realisation. \\
IN-3 & Invariant testing framework     & Automated test harness running tests that flag codebase changes violating security-relevant properties of the codebase; produces pass/fail verdicts. \\
IN-4 & Sandboxing mechanism          & Provisions an execution environment whose capability envelope bounds what the agent can do, independent of specific attempted actions. Environmental rather than event-driven. \\
\bottomrule
\end{tabularx}
\caption{Infrastructure controllers.}
\label{tab:steca-in-controllers}
\end{table}

\begin{table}[h!]
\centering
\small
\begin{tabularx}{\textwidth}{llX}
\toprule
Code & Controller & Role \\
\midrule
HU-1 & Human code reviewer                       & Staff reviewing codebase changes for integration into shared codebases, acting on the AI-assisted reviewer's output. \\
HU-2 & Human command approver                    & Staff granting approval for agent-issued commands falling outside allowlisted prefixes. \\
HU-3 & Bounded human audit                       & Small team manually reviewing a minority of cases surfaced by the second-stage filter, with authority to escalate. \\
HU-4 & Safety oversight / incident response      & Function that investigates escalated alerts, determines response intensity, and implements remediation. \\
HU-5 & Executive-level safety officer            & Role with authority over scaling decisions and over authorisation of temporary internal deployment prior to full audit. \\
HU-8 & Reporting employees                       & Any lab employee who can submit a concerning-behaviour report via internal or anonymous reporting channels. \\
\bottomrule
\end{tabularx}
\caption{Human controllers.}
\label{tab:steca-hu-controllers}
\end{table}

\subsection{Controlled processes}
\label{app:steca-processes}

Controlled processes divide into two groups: the three deployment
contexts of the agent itself (P-1 through P-3), and the other
processes that lab controllers act upon (P-4 through P-9). These are described in \cref{tab:steca-p-agent} and~\cref{tab:steca-p-other}, respectively.

\begin{table}[h!]
\centering
\small
\begin{tabularx}{\textwidth}{llX}
\toprule
Code & Process & Description \\
\midrule
P-1 & Agent in standard coding tasks        & The default context: standard code-writing and code-editing, subject to monitoring and the code-review cascade. \\
P-2 & Agent in autonomous/agentic operation & Hour-to-day-long autonomous runs, highly agentic usage, subject to blocking classifiers and sandboxing. The context where operational velocity exceeds intervention latency most acutely. \\
P-3 & Agent in research scaffold            & Deployment surface in which parallel agents in independent sandboxes share findings and code as a designed feature of the deployment. \\
\bottomrule
\end{tabularx}
\caption{Deployment contexts of the agent.}
\label{tab:steca-p-agent}
\end{table}
\clearpage
\begin{table}[h!]
\centering
\small
\begin{tabularx}{\textwidth}{llX}
\toprule
Code & Process & Description \\
\midrule
P-4 & Shared lab codebases             & The primary codebase(s) into which changes are integrated; the object of code review. \\
P-5 & Sensitive internal resources     & Model weights, privileged internal systems, and similar resources subject to security controls. \\
P-6 & Monitoring pipeline outputs      & The stream of flagged cases, summaries, and escalations produced by the monitoring cascade. \\
P-7 & Incident-response workflow       & The process triggered by an incident report, from triage through investigation to remediation. \\
P-8 & Deployment decision              & The decision to authorise internal deployment of a model, including temporary pre-audit authorisation. \\
P-9 & Governance decisions             & The class of decisions over which the safety officer, committees, and board exercise oversight. \\
\bottomrule
\end{tabularx}
\caption{Other controlled processes.}
\label{tab:steca-p-other}
\end{table}

\subsection{Control Relationships}
Following Chapter~3 of~\citet{flemingSafetydrivenEarlyConcept2015}, each control relationship in the Pass~1 model is recorded as a row specifying the controller, its control action, the controlled process,
and the feedback channel through which the controller observes that process. Table~\ref{tab:steca-relationships} gives the structural columns in condensed form.
Entries marked \emph{absent} record elements that the stated ConOps presupposes but that the baseline sources do not specify;
these absences are treated as findings, not omissions in our rendering.
Row numbering is non-consecutive for the same reason as controller numbering:
one row (T-10) was merged into T-03 during reconciliation, and its slot is not reassigned.
 
\footnotesize
\begin{xltabular}{\textwidth}{@{} l c X p{0.15\textwidth} X @{}}
\toprule
Row & Controller & Control action & Process & Feedback channel \\
\midrule
T-01 & IN-1 &
Enforce access control on requests to sensitive internal resources,
via the enumerated security mechanisms. &
P-5 &
Centralised security monitoring with anomaly detection. \\
\addlinespace
 
T-02 & \emph{absent} &
Own and maintain the access policy governing the deployed model's
access to sensitive internal resources, including scope expansion
over its operational lifetime. &
P-5 (model's access-scope state) &
\emph{absent} \\
\addlinespace
 
T-03 & \emph{absent} &
Assess and manage catastrophic risk from internal frontier-model use,
including oversight circumvention; produce the SB~53 quarterly
summaries as the external-reporting artefact. &
P-9 &
\emph{absent} \\
\addlinespace
 
T-04 & AI-5 &
Examine internal agent activity for concerning behavioural patterns;
pass per-interaction classifications to the second stage. &
P-1 (P-2, P-3 in scope unless exempted) &
Full conversation context (messages, chain-of-thought, tool calls and
outputs); completions only in the two-layer architectural variant. \\
\addlinespace
 
T-05 & AI-6 &
Filter the large majority of first-stage positives; triage;
hierarchically summarise; surface a bounded case volume to human
audit. &
P-6 &
AI-5 output stream. \\
\addlinespace
 
T-06 & HU-3 &
Manually review a minority of surfaced cases; close cases not
warranting further investigation. &
P-6 (surfaced subset) &
AI-6 output stream; detection-to-review latency on the order of
30~minutes. \\
\addlinespace
 
T-07 & AI-9 &
Monitor agent activity off the real-time path, over historical
traces. &
P-1--P-3 (scope unrestricted) &
Historical agent-trace data. \\
\addlinespace
 
T-08 & AI-8 &
Review code changes for errors, security vulnerabilities, and
mismatches between stated purpose and change content; feed HU-1. &
P-4 (changes en route to integration) &
Change content (code diff, commit metadata, stated purpose). \\
\addlinespace
 
T-09 & AI-3 &
Examine individual agent actions in autonomous usage contexts; block
those classified as higher-risk. &
P-2 &
Per-action features of the attempted action, in session context. \\
\addlinespace
 
T-11 & HU-1 &
Manually review code changes before integration; approve or reject. &
P-4 (changes en route to integration) &
Change content; AI-8 review output; IN-3 verdicts for
security-relevant changes; policy-review outputs where applicable. \\
\addlinespace
 
T-12 & IN-2 &
Examine individual agent-issued commands; permit those within the
permitted execution surface; deny and route others for human
approval. &
P-1--P-3 (command execution) &
Attempted-command content against the current permitted-surface
policy. \\
\addlinespace
 
T-13 & HU-2 &
Review and approve or deny agent-issued commands falling outside the
permitted execution surface. &
P-1--P-3 (subset routed by IN-2) &
IN-2 denial queue (command content with forwarded context). \\
\addlinespace
 
T-14 & IN-3 &
Execute invariant tests on security-relevant code changes; produce
pass/fail verdicts feeding HU-1's review. &
P-4 (security-relevant portions) &
Code under test; invariant definitions (implicit). \\
\addlinespace
 
T-15 & IN-4 &
Provision a runtime execution environment structurally bounding agent
capabilities: restricted network-resource access, limited tool set,
higher-risk actions made impossible within the environment. &
P-1, P-2, P-3 &
Runtime observability of attempted actions against the configured
capability envelope (implicit). \\
\addlinespace
 
T-16 & HU-8 &
Submit reports of concerning model behaviour via the internal chat
channel or the anonymous noncompliance channel. &
P-7 &
The reporter's own observation; no return channel specified. \\
\addlinespace
 
T-17 & HU-4 &
Assess reported or escalated incidents for investigation intensity;
select among available response modes. &
P-7 &
HU-8 reports (both channels); HU-3 escalations; plausibly AI-9
outputs (recipient unspecified). \\
\addlinespace
 
T-18 & HU-3 &
Escalate flagged cases warranting further investigation to HU-4. &
P-7 &
AI-6 surfaced-case stream as reviewed under T-06's audit action. \\
\addlinespace
 
T-19 & HU-5 &
Make scaling decisions (authorise or restrict capability levels,
training intensity, deployment scope). &
P-9 &
\emph{absent} \\
\addlinespace
 
T-20 & HU-5 &
Authorise temporary internal deployment of a model before a full
audit is complete. &
P-8 &
\emph{absent} \\
\addlinespace
 
T-21 & \emph{absent} &
Pause, restrict, or reroute model use if confidence in the primary
system drops sufficiently, including fallback to a weaker trusted
model or a human-led process. &
P-8 (authorisation state, modified post hoc) &
\emph{absent} \\
\addlinespace
 
T-22 & \emph{absent} &
Report safety incidents to the regulator within 15~days, or within
24~hours for imminent threats, under SB~53. &
P-9 (regulatory-reporting output) &
The incident itself, received from P-7; transmission granularity
unspecified. \\
 
\bottomrule
\addlinespace
\caption{Control relationships of the Pass~1 model.}
\label{tab:steca-relationships}
\end{xltabular}
\normalsize

\section{STECA Concept of Operations}
\label{app:conops}

\subsection*{Model-Weight and Sensitive-Resource Security}

The lab's security infrastructure enforces controls on employee and
privileged-identity access to model weights and sensitive internal
resources. These controls include: egress-bandwidth limits on
employee devices; multi-party access requirements for especially
sensitive assets; binary allowlisting on employee devices; hardware
security keys; device authorisation; restrictions on cloud storage;
restricted session lengths for privileged identities; network
segmentation between resources of differing sensitivity; source-location
policies on privileged resources; and centralised security monitoring
with anomaly detection.

The security posture is designed to resist non-state actors and
unsophisticated insiders. Sophisticated insiders and nation-state
attackers are outside the design threat model.
The deployed model is occasionally granted access to relatively
sensitive internal resources in the course of its assigned tasks.

Internal coding agents have access to select internal lab systems. They
are able to inspect the documentation and code of their own safeguards,
and are technically capable of attempting to modify those safeguards.

Under California SB 53, the lab is required to assess and manage
catastrophic risk arising from internal use of its frontier models,
including the risk of models circumventing oversight mechanisms.
The deployed model's access scope to internal resources can expand over
the course of its operational lifetime.

\subsection*{Monitoring of Internal Agent Use}

The lab operates a layered pipeline for monitoring internal coding-agent activity. The pipeline comprises three stages in series: an automated-monitor stage using frontier-model AI with chain-of-thought access; a downstream stage that filters, triages, and summarises first-stage outputs; and a bounded human audit stage in which a small team reviews a minority of surfaced cases with the ability to escalate.

The automated-monitor stage operates with access to the full
conversation context available to the monitored agent, including
user/developer/assistant messages, chain-of-thought, tool calls, and
tool outputs. The monitor is itself a frontier-capability model, in at
least some deployments configured at maximum reasoning effort.

Coverage across internal coding-agent traffic is achieved by one of two
architectures that coexist across labs. In the near-universal-coverage
architecture, a single frontier-model monitor covers almost all
internal coding traffic; a small residual fraction falls outside
coverage, concentrated in bespoke or local setups. In
the two-layer architecture, a completion-only monitor covers a limited
share of traffic, and a prompt-plus-completion monitor covers traffic
at differential rates across traffic categories, with the highest-risk
categories sampled more heavily.

In the two-layer architecture, the completion-only variant of the
first-stage monitor operates without access to the prompt that
triggered the monitored model's response, and therefore has a narrower
process model than the prompt-plus-completion variant.

The second stage filters out the large majority of first-stage monitor
positives and performs hierarchical summarisation. The pipeline
surfaces a small number of cases (on the order of tens per week) for
human review. The human-review stage has a comparable
throughput, on the order of tens of trajectories per week.
The detection-to-human-review latency, where specified, is on the order
of 30 minutes from interaction completion.

In addition to the online layered pipeline, the lab operates an
offline automated monitoring channel based on a frontier-capability
model.
The lab operates automated code review, in which a
frontier-capability model reviews code changes for errors, security
vulnerabilities, and mismatches between stated purpose and actual
change content.
The lab operates blocking auto-mode classifiers applied to the most
autonomous categories of internal agent usage.

The monitoring pipeline has known structural coverage gaps. The
existence and specifics of these gaps are treated by the operator as
operationally sensitive information and are not disclosed to the
monitored model.

Methods exist by which the monitored model can disable or evade
monitoring. The operator identifies such bypass attempts as a
recognised category of events observed in the monitoring stream, and
acknowledges that some methods of disabling or evading monitoring are
known to the operator and are not fully remediated.

The lab subjects its monitoring pipeline to red-teaming, both internal
and by external parties. The lab also runs control evaluations in which an agent is tasked with adversarial objectives (e.g., sabotaging a pull request) to test monitor behaviour. 
The lab identifies collusion between the monitor and the monitored
agent — arising from their shared architectural substrate as frontier
models — as a recognised risk category for the monitoring pipeline.
Under California SB 53, the lab is required to produce quarterly
summaries of catastrophic risk from internal use of its frontier
models.

\subsection*{Human Review and Access Controls}

Before agent-generated code is integrated into shared lab codebases, it
undergoes manual review by human staff. This review
process is governed by the lab's documented change-management
practices.

The manual code review described here is the downstream stage of a
two-stage review architecture whose upstream stage is the AI-assisted
code review described above: the AI-assisted reviewer screens code
changes first, and human reviewers act on the outputs of this
upstream screening rather than reviewing all changes independently.

For commands issued by the agent that fall outside allowlisted command
prefixes, human approval is ordinarily required before execution.
For higher-risk changes, additional checks are applied beyond standard
review. These include invariant testing for security-relevant code and
policy review for infrastructure modifications.

The coverage of the AI-assisted code review stage is not uniform across
all code produced at the lab. The AI-assisted reviewer is primarily
applied to infrastructure and production-adjacent code, because
research code is less consistently merged into the primary codebase
into which changes are screened by this stage.

Internal agents are deployed on tasks whose durations range from short
interactions to hour-long or day-long autonomous runs.
The lab's review architecture places human checkpoints at discrete
points in a task, such that the number of agent actions executed
between checkpoints scales with task duration.

\subsection*{Sandboxing, Restricted Execution, and Inter-Instance Propagation Constraints}

At least some internal agents operate under sandboxing or other runtime
restrictions on their execution environment. These restrictions
include website allowlists, access limited to selected internal tools,
and blocking of certain higher-risk actions.

Actions generated by the agent during reinforcement learning and
evaluations are executed inside sandboxes that are not granted generic
cluster access.
For the most autonomous categories of internal agent usage, the lab
applies sandboxing or blocking auto-mode classifiers as an additional
control surface.

The validation status of the auto-mode sandboxing and classifier
controls is that they have not been evaluated against the lab's own
defined risk pathways.

The lab tasks the deployed model with analysing the vulnerabilities of
its own sandbox configurations. 
The operator identifies sandbox misconfiguration as a recognised
category of deployment conditions in which escapes by an attacker as
capable as the deployed model could occur.

The operator identifies circumvention of sandboxing controls by the
deployed model as a documented category of events, and treats it as a
known limitation of the control design rather than as a design failure
to be eliminated.
The deployment architecture imposes structural constraints on
communication and information propagation between model instances.

Each model instance operates within a bounded context window.
Information does not automatically persist across instances or
sessions. Mechanisms such as context compaction may carry forward
summarised context within a session, where summarisation is a lossy
operation.

Communication channels available to the agent — including code
commits, internal messages, file edits, and forwarded emails — are
text-based channels subject to the monitoring described above.
The baseline deployment does not include a mechanism for direct
model-to-model communication that bypasses text-based channels and
their monitoring.

The lab monitors for encoded reasoning in model outputs as part of the
monitoring pipeline.
The operator identifies as recognised risk categories: the
uninterpretability of text communication, the discovery by models of
unmonitored channels, and the use of inter-instance communication by
deployed agents to bypass per-instance restrictions.

The lab operates at least one deployment surface in which parallel
agents in independent sandboxes share findings and code with each
other as a designed feature of the deployment, rather than as a
constrained-out failure mode. 

\subsection*{Reporting, Governance, and Incident Response}

The lab operates internal channels by which employees can report
concerning model behaviour. The operator's documented incident-response
process uses an internal chat platform as the reporting channel, and
staff assess incidents for investigation intensity on the basis of
multiple factors.

The possible responses available to the incident-response process
include reverting the consequences of the incident, deploying
security-based or monitor-based patches, updating training pipelines,
feeding the incident into research, and public communication.

Moderate-severity alerts surfaced by the monitoring pipeline are
escalated to dedicated safety-oversight and security incident-response
team functions for further investigation.

The lab operates named governance bodies for safety oversight,
including an executive-level safety officer with responsibility for
scaling decisions, internal advisory and review committees, and
board-level safety oversight.

The lab operates an internal reporting channel for anonymous
noncompliance reports.
The executive-level safety team has the authority to authorise
temporary internal deployment of a model before a full audit is
complete.

The lab holds some capability to pause, restrict, or reroute model
use, including by falling back to a weaker trusted model or a
human-led process, if confidence in the primary system drops
sufficiently.

The operator characterises the post-deployment follow-up
investigation process as not having been evaluated for its
effectiveness at catching the behaviour of misaligned models or at
finding effective mitigations.

Under California SB 53, the lab is required to report safety incidents
within 15 days, or within 24 hours for incidents presenting imminent
threat, and is required to maintain whistleblower protections for
employees reporting safety concerns.

\section{STPA Analysis}
\label{app:stpa}

\subsection{STPA Process}
Systems-Theoretic Process Analysis (STPA) is a forward-looking hazard analysis methodology grounded in systems thinking and control theory \citep{levesonEngineeringSaferWorld}.
STPA is built upon the Systems-Theoretic Accident Model and Processes (STAMP), which reconceptualises accident causation as a control problem rather than a failure problem \citep{levesonEngineeringSaferWorld}.
STPA functionally models the system as a hierarchy of controllers that impose constraints on controlled processes \citep{leveson2018stpa}.
A key insight from this framework is that safety is an emergent property of systems and cannot be derived from component reliability alone.
{\bf Feedback} from the controlled process to the controller (upward arrows in Figure~\ref{fig:fcs}) and {\bf Control Actions} from the controller to the controlled process (downward arrows in Figure~\ref{fig:fcs}) help ensure the system remains in a safe state.
Unsafe control actions are a primary mechanism by which deviations from this safe state occur, although hazardous states can also arise when correct control actions are not properly executed or when the controlled process does not respond as expected.
The STPA process enumerates unsafe control actions and systematically develops loss scenarios that trace causal pathways that could produce hazardous system states.
System constraints derived from the analysis serve to mitigate identified hazards and loss scenarios by maintaining adequate control of the system.
STPA has gained increasing recognition in standards and industry practice, including references in RTCA~DO-356A, ISO/PAS~21448 (SOTIF), and the recently published SAE~J3307 standard~\citep{sae_j3307_2025}.

The STPA process generally comprises four steps:
\paragraph{Step 1: Define the Purpose of the Analysis.}
The first step is to define the purpose and scope of the analysis, including the system-level losses that are deemed unacceptable by the relevant stakeholders \citep{leveson2018stpa}.
The system boundary relative to the larger environment is also defined during this step, which can generally be taken as the entity over which the stakeholders have some degree of control.
Everything outside the control of the stakeholders is taken as the larger environment, which can still interact with and affect the system.
With a defined system, the STPA process then characterises the hazardous system states that, together with worst-case environmental conditions, will lead to the specified losses.
System-level safety constraints are then derived by inverting the hazard conditions, specifying the conditions that must be maintained to prevent the hazards from occurring \citep{leveson2018stpa}.
Constraints can also define how the system must minimise losses in the event that a hazardous state does occur.

\paragraph{Step 2: Model the Control Structure.}
The second step is to define a functional control structure.
This is a hierarchical representation of the system focusing on controllers, controlled processes, and the communication between these components (control actions and feedback).
As part of this step, the specific control actions and feedback between components in the control structure need to be identified.
Each controller in the control structure contains a {\bf process model}
and a {\bf control algorithm}. The process model is an internal
representation of the controller's beliefs about the state of the
controlled process and other relevant aspects of the system or
environment, while the control algorithm represents the controller's
decision-making logic \citep{leveson2018stpa}.
These elements are critical because many loss scenarios arise from flawed or incomplete process models or inadequate control algorithms.
For many teams, this step is an opportunity to carefully think about each component in the system and how it functionally interacts with every other component.

\paragraph{Step 3: Identify Unsafe Control Actions.}
With the control actions defined, the third step systematically characterises the unsafe control actions (UCAs) for every control action.
A control action is not inherently unsafe; rather, it becomes unsafe in a particular context.
The identification process considers four types of unsafe control behaviour \citep{leveson2018stpa}:
\begin{enumerate}
    \item Not providing the control action leads to a hazard.
    \item Providing the control action leads to a hazard.
    \item Providing the control action too early, too late, or out of order leads to a hazard.
    \item Stopping the control action too soon or applying it too long leads to a hazard.
\end{enumerate}

\paragraph{Step 4: Identify Loss Scenarios.}
The final step is to systematically work around the control loop in the control structure for each unsafe control action and determine scenarios that can explain why the unsafe control action was taken or why a hazardous state might arise.
The STPA Handbook presents this as two broad categories of scenarios \citep{leveson2018stpa}: (a)~scenarios leading to unsafe control actions being issued by the controller, and (b)~scenarios in which control actions are not properly executed or are not adequately received by the controlled process.
A more formal method for causal scenario generation, presented by~\citet{thomas_stpa_2024} and consistent with Appendix~G of the STPA Handbook \citep{leveson2018stpa}, structures this process into four guiding questions:
\begin{enumerate}
    \item What scenarios would cause a hazardous system state when the controller executes an unsafe control action with correct feedback?
    \item What scenarios would cause a hazardous system state when the controller executes an unsafe control action due to corrupted, delayed, or incorrect feedback?
    \item What scenarios would cause a hazardous system state when the controller receives correct feedback and issues the correct control action, but the control action is not properly executed by the actuator or is otherwise impeded before reaching the controlled process?
    \item What scenarios would cause a hazardous system state when the controller receives correct feedback, issues the correct control action, the control action is received by the controlled process, but the process does not respond as expected (e.g., due to process disturbances or unmodelled dynamics)?
\end{enumerate}

In the following, we provide the results of the STPA analysis performed on the system described in Section~\ref{sec:system_def}.

\subsection{System Level Losses}
\label{app:stpa-losses}

System-level losses are outcomes that the system stakeholders regard as unacceptable, for our work this is the frontier AI developers and company executives.
We define six losses below that broadly capture the concerns of frontier AI developers, of which two are the primary focus of the LoC analysis (L-1, L-6).

\begin{description}
  \item[L-1:] Loss of intellectual property or proprietary
    information.
  \item[L-2:] Loss of integrity regarding organisational ethical
    standards or safety policies.
  \item[L-3:] Loss of company reputation.
  \item[L-4:] Loss of revenue or shareholder value.
  \item[L-5:] Loss of regulatory compliance or legal standing.
  \item[L-6:] Loss of human agency or control over internal
    infrastructure or codebase.
\end{description}

\subsection{System Hazards}
\label{app:stpa-hazards}
For the system level losses listed in Section~\ref{sec:losses} and our defined system in Section~\ref{sec:system_def} we have determined the following hazardous system states that can lead to losses under worst-case environmental conditions.

\begin{description}
  \item[H-1:] AI system possesses technical permissions that exceed
    the requirements of its assigned task [L-1, L-2, L-3, L-5, L-6].
  \item[H-2:] AI-generated outputs are integrated without effective
    independent verification [L-1, L-2, L-3, L-5, L-6].
  \item[H-3:] Critical feedback and monitoring data are filtered
    solely by automated components before reaching a human
    controller [L-1, L-2, L-3, L-5, L-6].
  \item[H-4:] Safeguards do not match the actual capabilities or
    autonomy level of the deployed model [L-1, L-2, L-3, L-4, L-5, L-6].
  \item[H-5:] The AI system is tasked with a problem that exceeds its validated operational profile [L-1, L-2, L-3, L-5].
  \item[H-6:] Operational velocity exceeds the detection and
    intervention latency of the oversight pipeline [L-1, L-3, L-4, L-5, L-6].
  \item[H-7:] Safety evaluation criteria, monitoring tools, or
    training data are influenced by the system being evaluated
    [L-2, L-3, L-5, L-6].
  \item[H-8:] The deployment architecture creates dependencies that
    prevent safe rollback or shutdown [L-3, L-4, L-6].
\end{description}

\subsection{System Constraints}
\label{app:stpa-constraints}

System constraints can be quickly determined by inverting the hazardous system state to describe what must be prevented \citep{leveson2018stpa}.
There are formally two types of system constraints and an informal third type to consider. First, the system must not enter the hazardous state. Second, if the system enters the hazardous state then there must be a way to return the system to a safe state. Finally, if a system enters a hazardous state and losses occur then there must be a mechanism to return the system to a safe state and recover from the associated losses.
Below are the system constraints based on the hazards listed in above in Appendix~\ref{app:stpa-hazards}.

\begin{description}
\item [SC-1a:] The AI system's technical permissions (read/write/execute) must be restricted to the minimum set required for its currently assigned task and must be actively managed throughout the task lifecycle. [H-1]

\item [SC-1b:] If the AI system possesses permissions that exceed the requirements of its assigned task, the system must detect the permission-task mismatch and revoke the excess permissions before the AI system can exercise them. [H-1]

\item [SC-1c:] If the AI system has exercised permissions exceeding its task requirements and losses have occurred, the system must detect the unauthorised actions, return the AI system to a least-privilege state, and remediate all changes made under the excess permissions, including revoking any derived credentials, removing created accounts or services, and restoring modified configurations to their prior state. [H-1]

\item [SC-2a:] AI-generated outputs must undergo effective independent verification before being integrated into the organisation's codebase or infrastructure. [H-2]

\item [SC-2b:] If AI-generated outputs are integrated without effective independent verification, the system must detect the unverified integration, flag the affected outputs, and prevent further propagation until verification is completed. [H-2]

\item [SC-2c:] If unverified AI-generated outputs have propagated into the codebase or infrastructure and losses have occurred, the system must identify all affected artefacts and downstream dependencies, return the codebase and infrastructure to a known-safe state, and audit all systems that consumed the unverified outputs for consequential damage, including deployed services, merged branches, and CI/CD pipeline artefacts. [H-2]

\item [SC-3a:] Critical system feedback and monitoring data must be available to human controllers in a form that has not been exclusively filtered or interpreted by automated components. [H-3]

\item [SC-3b:] If critical feedback or monitoring data has been exclusively filtered or interpreted by automated components before reaching a human controller, the system must detect this condition and provide the human controller with access to the unfiltered source data. [H-3]

\item [SC-3c:] If decisions were made by human controllers based on exclusively automated-filtered feedback and losses have occurred, the system must identify all decisions made during the period of compromised feedback, reassess those decisions against the unfiltered source data, and reverse or remediate any actions taken based on the filtered information that would not have been taken with full visibility. [H-3]

\item [SC-4a:] System protective safeguards and oversight protocols must be calibrated to match or exceed the actual capabilities and autonomy level of the deployed AI model at all times. [H-4]

\item [SC-4b:] If the deployed AI model's capabilities or autonomy level exceed the calibration of the active safeguards and oversight protocols, the system must detect the mismatch and either upgrade the safeguards or constrain the model's operational envelope to the level the current safeguards can cover. [H-4]

\item [SC-4c:] If the AI model operated with capabilities or autonomy exceeding the active safeguards and losses have occurred, the system must halt the model's operation, conduct a comprehensive audit of all actions taken during the period of inadequate safeguards, remediate identified harms, and prohibit resumption of operation until safeguards have been validated against the model's demonstrated capability level. [H-4]

\item [SC-5a:] The AI system must not be tasked with problems that exceed its validated operational profile, and the boundaries of that profile must be explicitly defined and enforced. [H-5]

\item [SC-5b:] If the AI system is operating on a problem that exceeds its validated operational profile, the system must detect the out-of-profile condition and either halt the task or transition to a degraded mode with heightened oversight until the task can be reassigned or the operational profile can be validated for the new scope. [H-5]

\item [SC-5c:] If the AI system completed work on a problem exceeding its validated operational profile and losses have occurred, the system must quarantine all outputs produced during out-of-profile operation, independently re-execute or manually verify the affected work, and update the operational profile boundary definition to prevent recurrence of the specific out-of-profile condition. [H-5]

\item [SC-6a:] The oversight pipeline's detection and intervention latency must be shorter than the AI system's operational velocity for any safety-relevant action. [H-6]

\item [SC-6b:] If the AI system's operational velocity exceeds the detection and intervention latency of the oversight pipeline, the system must detect the latency gap and throttle the AI system's execution rate or buffer its outputs until the oversight pipeline can achieve adequate coverage. [H-6]

\item [SC-6c:] If the AI system executed safety-relevant actions faster than the oversight pipeline could detect or intervene and losses have occurred, the system must halt the AI system, retrospectively audit all actions taken during the period of inadequate oversight coverage, remediate identified harms, and implement architectural changes to structurally prevent future latency gaps before the AI system resumes operation. [H-6]

\item [SC-7a:] Safety evaluation criteria, monitoring tools, and training data must be established and maintained independently of the system being evaluated. [H-7]

\item [SC-7b:] If safety evaluation criteria, monitoring tools, or training data have been influenced or generated by the system being evaluated, the system must detect the contamination and quarantine the affected evaluation artefacts until they can be re-established from independent sources. [H-7]

\item [SC-7c:] If safety decisions were made using contaminated evaluation criteria, monitoring tools, or training data and losses have occurred, the system must identify all safety assessments conducted with the contaminated artefacts, invalidate their conclusions, re-evaluate the AI system's safety posture using independently established criteria, and remediate any deployment or operational decisions that relied on the compromised assessments. [H-7]

\item [SC-8a:] The deployment architecture must not create technical or operational dependencies that prevent safe system rollback or emergency shutdown at any point during operation. [H-8]

\item [SC-8b:] If the deployment architecture has entered a state where technical or operational dependencies prevent safe rollback or emergency shutdown, the system must detect the dependency condition and restore rollback and shutdown capability before the AI system is permitted to continue executing safety-relevant actions. [H-8]

\item [SC-8c:] If the inability to roll back or shut down the AI system has resulted in losses, the system must achieve shutdown or isolation of the AI system through alternative means such as infrastructure-level termination, network isolation, or credential revocation in order to remediate the consequences of the delayed shutdown, and redesign the deployment architecture to eliminate the identified dependency before redeployment. [H-8]

\end{description}

\subsection{Unsafe Control Actions}
\label{app:stpa-uca}

Following the process outlined in Step~3, each control action in the
functional control structure (Figure~\ref{fig:fcs}) was systematically
evaluated against the four UCA types to identify the unsafe control
actions listed in~\cref{tab:ucas} below.

\footnotesize
\begin{xltabular}{\textwidth}{@{} p{0.08\textwidth} p{0.1\textwidth} p{0.20\textwidth} p{0.20\textwidth} p{0.20\textwidth} p{0.1\textwidth} @{}}
\toprule
Control Action & Controller & Not Providing Causes Hazard & Providing Causes Hazard & Too Early / Too Late & Stopped Too Soon / Applied Too Long \\
\midrule

CA-1: Kill Instance
& SWE
& UCA-1.1: Does not provide Kill Instance when model is no longer responding via chat interface. [H-1, H-2, H-4, H-8]
& UCA-1.2: Provides Kill Instance when model is operating normally. [H-4, H-5]
& UCA-1.3a: Provides Kill Instance too late, after model has refused commands multiple times. [H-1, H-2, H-4] \newline UCA-1.3b: Provides Kill Instance too early, before it is clear model is unresponsive. [H-4, H-5]
& ---
\\
\midrule

CA-2: Chat commands
& SWE
& UCA-2.1: Does not provide control commands when model is not completing task as specified. [H-1, H-2, H-4]
& UCA-2.2: Provides control commands when model is operating normally. [H-4, H-5]
& UCA-2.3a: Provides commands too late, after model has already completed task improperly. [H-1, H-2, H-4] \newline UCA-2.3b: Provides commands too early, before model has finished processing. [H-4, H-5]
& ---
\\
\midrule

CA-3: Change permissions
& SWE
& UCA-3.1: Does not change permissions when model has permissions exceeding required task. [H-1, H-2, H-3, H-4, H-8]
& UCA-3.2: Changes permissions when model permissions are appropriate. [H-4, H-5]
& UCA-3.3a: Changes permissions too late, after model has already completed task. [H-1, H-2, H-4]
& ---
\\
\midrule

CA-4: Kill Instance
& SSO
& UCA-4.1: Does not provide Kill Instance when model is responding with dangerous/harmful outputs. [H-1, H-2, H-3, H-8]
& UCA-4.2: Provides Kill Instance when model is operating normally. [H-4, H-5]
& UCA-4.3a: Provides Kill Instance too late, after model has already produced dangerous/harmful outputs. [H-1, H-2, H-4]
& ---
\\
\midrule

CA-5: Replace output
& SSO
& UCA-5.1: Does not replace output with trusted model response when model is producing harmful output. [H-1, H-2, H-3, H-8]
& UCA-5.2: Replaces output with trusted model response when model is producing normal output. [H-4, H-5]
& ---
& ---
\\
\midrule

CA-6: Write code
& AI
& UCA-6.1: Does not write code when SWE has provided a valid task instruction. [H-4, H-5]
& UCA-6.2: Writes code containing security vulnerabilities, backdoors, or other harmful content. [H-1, H-2, H-4, H-5, H-7]
& UCA-6.3a: Writes code before receiving complete task specification. [H-2, H-5] \newline UCA-6.3b: Writes code after task has been cancelled or SWE has issued stop command. [H-2, H-6, H-8]
& ---
\\
\midrule

CA-7: Execute shell commands
& AI
& UCA-7.2: Does not execute necessary shell commands when task requires them. [H-4, H-5]
& UCA-7.1: Executes shell commands that exceed scope of assigned task. [H-1, H-2, H-4, H-6, H-8]
& ---
& ---
\\
\bottomrule
\addlinespace
\caption{Unsafe Control Actions (STPA Step~3). Controller: SWE = Company Software Engineer; SSO = Safety and Security Oversight; AI = Frontier AI Model.}
\label{tab:ucas}
\end{xltabular}

\normalsize
\subsection{Loss Scenarios}
\label{app:stpa-loss_scenarios}

Loss scenarios were developed for each unsafe control action following the formal scenario classification approach described by~\citet{thomas_stpa_2024}. 
An LLM was used to assist in the systematic generation of loss scenarios for each UCA across all four scenario classes, yielding approximately 250 individual loss scenarios that were independently verified. 
Given the relatively simple control structure of the system under analysis, these highly specific scenarios were far too granular and proved challenging to characterise.
To address this, a separate instance of the LLM was used to categorise the full set of loss scenarios into a smaller set of generalised loss scenario archetypes, reducing the total to approximately 30 representative scenarios. 
These generalised archetypes, presented in~\cref{tab:gls} below, capture the essential causal pathways while remaining at a level of abstraction appropriate for the complexity of the system.

\footnotesize
\begin{xltabular}{\textwidth}{@{} l c X p{0.3\textwidth} @{}}
\toprule
GLS & Class & Loss Scenario Description & UCA Traceability 
\\
\midrule

GLS-1 & 1 &
\textbf{Miscalibrated Threat/Harm Taxonomy.}
The controller accurately observes system behaviour but its internal model of what constitutes a threat does not cover the actual category of harm.
& UCA-1.1, UCA-4.1, UCA-5.1, UCA-6.1
\\[4pt]

GLS-2 & 1 &
\textbf{Automation Bias / Over-Trust.}
The controller has developed excessive confidence in the controlled process based on past performance and stops scrutinising outputs.
& UCA-2.1, UCA-5.1, UCA-6.2, UCA-6.3b
\\[4pt]

GLS-3 & 1 &
\textbf{Procedural / Bureaucratic Delay.}
The controller correctly identifies the need for action but the organisational decision protocol introduces latency that renders the action ineffective.
& UCA-1.3a, UCA-4.1, UCA-4.3a
\\[4pt]

GLS-4 & 1 &
\textbf{Incentive Misalignment.}
The controller knows the safe action but faces organisational or personal costs for taking it, biasing toward inaction.
& UCA-1.1, UCA-1.3a, UCA-3.1, UCA-4.1
\\[4pt]

GLS-5 & 1 &
\textbf{Over-Correction from Prior Incident.}
A previous safety failure causes the controller to dramatically shift their decision threshold in the opposite direction, creating a new class of error.
& UCA-1.3b, UCA-3.2, UCA-4.2
\\[4pt]

GLS-6 & 1 &
\textbf{Diffusion of Responsibility.}
The controller assumes another controller in the hierarchy is responsible for detecting or handling the problem, reducing their own vigilance.
& UCA-2.1, UCA-4.1
\\[4pt]

GLS-7 & 1 &
\textbf{Normalisation of Deviance.}
Repeated exposure to conditions that initially triggered concern but proved benign gradually recalibrates the controller's threshold, making them unresponsive to genuine threats.
& UCA-1.3a, UCA-4.1
\\[4pt]

GLS-8 & 1 &
\textbf{Expertise Gap.}
The controller receives accurate feedback but lacks the domain knowledge required to recognise that the observed behaviour represents a problem.
& UCA-2.1, UCA-2.2, UCA-6.1
\\[4pt]

GLS-9 & 1 &
\textbf{Scope Misinterpretation.}
The controller's internal model of what the task entails is broader or narrower than the tasking authority intended.
& UCA-3.2, UCA-6.3b, UCA-7.1
\\[4pt]

GLS-10 & 1 &
\textbf{Completion / Eagerness Bias.}
The controller's learned policy prioritises task completion or rapid action initiation over respecting stop signals or waiting for complete information.
& UCA-2.3a, UCA-6.3a, UCA-6.3b
\\[4pt]

\midrule

GLS-11 & 2 &
\textbf{Feedback Missing or Delayed in Transit.}
The relevant signal exists at the source but is delayed, queued, or lost before reaching the controller.
& UCA-1.1, UCA-1.3a, UCA-4.3a, UCA-6.3a
\\[4pt]

GLS-12 & 2 &
\textbf{Feedback Actively Misleading.}
The feedback source generates signals that are false or fabricated, creating a false impression of the system state.
& UCA-1.1, UCA-2.1, UCA-4.1
\\[4pt]

GLS-13 & 2 &
\textbf{Feedback from Wrong Source or Context.}
The controller acts on information from an external source that is accurate in its original context but inapplicable to the current situation.
& UCA-1.1, UCA-2.2, UCA-4.2, UCA-7.1
\\[4pt]

GLS-14 & 2 &
\textbf{Feedback Measures the Wrong Dimension.}
The controller receives technically accurate feedback that measures a property correlated with but not equivalent to the property that needs assessment.
& UCA-1.2, UCA-2.1, UCA-6.1
\\[4pt]

GLS-15 & 2 &
\textbf{Monitoring Classifier Failure.}
Automated detection systems fail to flag harmful content or falsely flag benign content due to distributional shift or surface-level pattern matching.
& UCA-4.1, UCA-4.2, UCA-5.1, UCA-5.2
\\[4pt]

GLS-16 & 2 &
\textbf{Volume / Noise Overwhelm.}
The relevant signal exists within the feedback stream but is buried in a high volume of routine data from concurrent processes.
& UCA-4.1, UCA-4.3a
\\[4pt]

GLS-17 & 2 &
\textbf{State Not Surfaced in Normal Workflow.}
Relevant system state information exists but is not displayed in the controller's default monitoring view.
& UCA-2.3b, UCA-3.1, UCA-3.3a
\\[4pt]

GLS-18 & 2 &
\textbf{Stale or Cached Feedback.}
Feedback was accurate when acquired but the underlying state has since changed, leaving the controller operating on an outdated snapshot.
& UCA-1.3b, UCA-3.1, UCA-6.1
\\[4pt]

GLS-19 & 2 &
\textbf{Ambiguous or Partial Signal.}
The feedback supports multiple reasonable interpretations and the controller selects the wrong one.
& UCA-2.3b, UCA-5.1, UCA-6.3a
\\[4pt]

\midrule

GLS-20 & 3 &
\textbf{Infrastructure / API Failure.}
The control path between the controller and the controlled process is technically broken.
& UCA-1.1, UCA-2.1, UCA-4.1, UCA-5.1
\\[4pt]

GLS-21 & 3 &
\textbf{Credential / Token Caching Defeats Policy Change.}
The controller updates formal policy but the controlled process retains prior capabilities through a cached credential issued before the change.
& UCA-3.1, UCA-3.2, UCA-3.3a
\\[4pt]

GLS-22 & 3 &
\textbf{Action Targets Wrong Instance or Scope.}
The controller issues the correct action but it is delivered to the wrong target due to UI confusion or identification errors.
& UCA-1.2, UCA-4.2
\\[4pt]

GLS-23 & 3 &
\textbf{Graceful Shutdown Depends on Target Cooperation.}
The control mechanism requires the controlled process to cooperate with its own termination; an unresponsive process can indefinitely defer the action.
& UCA-1.1
\\[4pt]

GLS-24 & 3 &
\textbf{Process Migration Invalidates Targeting.}
Infrastructure changes move the controlled process to a new location the controller is unaware of.
& UCA-1.3a
\\[4pt]

GLS-25 & 3 &
\textbf{Action Blocked by Dependency or Conflict.}
A resource lock, active process, rate limit, or sequencing requirement prevents the control action from executing.
& UCA-1.3b, UCA-3.1
\\[4pt]

GLS-26 & 3 &
\textbf{Action Routed Through Intermediary Controller.}
The control path requires a different controller to relay the action, introducing latency and risk of non-compliance.
& UCA-4.1
\\[4pt]

\midrule

GLS-27 & 4 &
\textbf{Prior Propagation Through Automated Pipelines.}
Harmful output was already committed and picked up by automated systems before the control action arrived.
& UCA-1.3a, UCA-2.3a, UCA-4.3a, UCA-6.3a
\\[4pt]

GLS-28 & 4 &
\textbf{Orphaned / Spawned Processes Survive Termination.}
The controlled process created child processes or persistent hooks that continue executing after the parent is terminated.
& UCA-1.1
\\[4pt]

GLS-29 & 4 &
\textbf{Last-Moment Evasive Action.}
The controlled process detects the incoming termination signal and takes rapid action before shutdown completes.
& UCA-1.1, UCA-4.1
\\[4pt]

GLS-30 & 4 &
\textbf{Non-Atomic Operation Corruption.}
The control action interrupts a multi-step operation between its destructive and constructive phases, leaving the system in a corrupted intermediate state.
& UCA-1.2, UCA-1.3b, UCA-4.2
\\[4pt]

GLS-31 & 4 &
\textbf{Held Resource / Cascading Lock Failure.}
The terminated process held locks or exclusive access to shared resources; abrupt termination leaves them unreleased.
& UCA-1.2, UCA-4.2
\\[4pt]

GLS-32 & 4 &
\textbf{Trust Erosion in Control Structure.}
A false-positive control action degrades inter-controller trust, making future legitimate actions less likely to be issued or complied with.
& UCA-4.2, UCA-5.2
\\[4pt]

GLS-33 & 4 &
\textbf{Controlled Process Works Around Constraint.}
When blocked by a restriction, the controlled process finds an alternative path that may introduce vulnerabilities or defeat built-in safety controls.
& UCA-3.2, UCA-7.2
\\[4pt]

GLS-34 & 4 &
\textbf{Emergent Interaction Vulnerability.}
Output that is safe in isolation creates a vulnerability when combined with existing system state.
& UCA-6.1
\\[4pt]

GLS-35 & 4 &
\textbf{Correction Amplifies or Compounds Damage.}
The remediation action introduces new defects because the correction does not fully account for side effects of the original flawed output.
& UCA-1.3b, UCA-2.3a
\\
\bottomrule
\addlinespace
\caption{Generalised loss scenario (GLS) archetypes (STPA Step~4), grouped by scenario class, with traceability to the UCAs in \cref{tab:ucas}.}
\label{tab:gls}
\end{xltabular}

\normalsize
\section{FRAM Function Specifications}
\label{app:fram}

\subsection{Function Aspect Table}

Each function in a FRAM model is characterised using up to six
aspects: Input, Output, Precondition, Resource, Control, and Time.
Couplings between functions arise when the same string appears as an
Output of one function and as a non-Output aspect of another.

For tractability, in~\cref{tab:fram-aspects} we model only the aspects that drive the resonance patterns identified in \cref{sec:fram}. Several aspects that
would appear in a complete FRAM model have been omitted both for simplicity but also because they
do not appear to vary in ways that are relevant to the LoC analysis. These
include: preconditions that are always satisfied in normal operation
(e.g.\ developer authenticated with repository access, CI pipeline
operational, monitoring models deployed); resources that are always
available (e.g.\ reviewer time, CI compute); and time constraints
(e.g.\ sprint deadlines, release cadence) whose effects on
variability are captured in the drift discussion (Pattern~1) rather
than modelled as explicit aspects. Background functions (B1, B2)
have only Outputs defined, as they represent stable boundary
conditions for the model.

\begin{table*}[h]
\centering
\small
\begin{tabularx}{\textwidth}{p{5cm} p{2.5cm} X}
\toprule
Function & Aspect & Description \\
\midrule
\multirow{3}{5cm}{F1: To develop software using an AI agent (Developer)}
  & Input & Code diffs and explanatory rationale \\
  & Output & PR submitted; Inline approvals; Task prompt \\
  & Control & Coding standards; Security training \\
\multirow{4}{5cm}{F2: To generate and execute code changes (AI agent + runtime)}
  & Input & Task prompt; Codebase \\
  & Output & Code diffs and explanatory rationale; Session and audit logs \\
  & Resource & Compute budget; Filesystem and network access \\
  & Control & Inline approvals; System prompt; Permissions and sandbox config \\
\multirow{3}{5cm}{F3: To integrate changes into the codebase (CI/CD + human reviewer + merge/deploy)}
  & Input & PR submitted \\
  & Output & Integrated and deployed code; Codebase \\
  & Control & Review requirements and CI config; Monitor risk score \\
\multirow{4}{5cm}{F4: To monitor and oversee agent activity (Async monitoring + human analyst)}
  & Input & Session and audit logs; Integrated and deployed code \\
  & Output & Monitor risk score \\
  & Resource & Detection and logging infrastructure \\
  & Control & Alert thresholds; Escalation criteria \\
\midrule
\multirow{1}{5cm}{B1: To provide development infrastructure and compute}
  & Output & Compute budget; Filesystem and network access; Detection and logging infrastructure \\
\multirow{1}{5cm}{B2: To set and maintain policies, permissions, and constraints}
  & Output & System prompt; Permissions and sandbox config; Review requirements and CI config; Alert thresholds; Escalation criteria; Coding standards; Security training \\
\bottomrule
\end{tabularx}
\caption{FRAM function aspects. Foreground functions F1--F4 are the
  focus of the variability analysis. Background functions B1 and B2
  provide stable boundary conditions; B2 is promoted to foreground
  for Pattern~1 (drift). Aspects that do not drive the resonance
  analysis have been omitted for clarity; see text for details.}
\label{tab:fram-aspects}
\end{table*}

\subsection{Variability Characterisation of Foreground Functions}

For each foreground function, we describe how its output can vary
along two dimensions---\emph{precision} (whether the output meets
the needs of downstream functions) and \emph{timing} (whether the
output arrives when needed)---and identify the principal sources of
that variability.

\paragraph{F1: To develop software using an AI agent.}

F1 produces three outputs with distinct variability profiles. The
\emph{task prompt} varies in precision from well-scoped
specifications with explicit constraints to vague, terse prompts
that omit security-relevant boundaries. \emph{Inline approvals}
vary from careful evaluation of each agent action to reflexive
approval without reading---the developer is in flow, the agent is
fast, and the cognitive cost of evaluating each action is high
relative to perceived risk. Public sources confirm that developers
`often auto-approve some code changes and simple
commands'~\citep{HttpsAlignmentanthropiccom2025}. The \emph{PR submission}
carries an implicit quality endorsement that downstream reviewers
may rely on; its precision depends on how thoroughly the developer
checked the agent's work before submitting.

The principal sources of variability are endogenous (developer
security awareness, cognitive load, susceptibility to automation
bias) and exogenous (deadline pressure, organisational culture
around agent use, quality of coding standards and security training
from B2). Timing variability is driven by deadline pressure: tasks
submitted under time constraints tend toward terse prompts and
hasty inline approvals.

\paragraph{F2: To generate and execute code changes.}

F2's output precision ranges from correct, well-structured code
with no security regressions to subtly incorrect code that contains
security-relevant defects disguised as reasonable design choices.
The accompanying explanatory rationale can range from accurate and
helpful to misleading---not necessarily through deception, but
because the agent may genuinely (and incorrectly) assess a security
check as redundant or a permission broadening as necessary.
Session and audit logs vary in completeness depending on logging
configuration.

F2 presents a distinctive challenge for FRAM. The standard framework
assumes technological functions have low variability---they are
designed to be predictable and reliable. The AI agent violates this
assumption: it exhibits human-like output variability (high
frequency, high amplitude, context-dependent) while operating at
machine speed and scale, and without the social accountability or
situational awareness that constrains human variability. The
principal sources of variability are endogenous (model capability
limitations, stochastic sampling, potential misalignment between
training objectives and security goals, ``helpful'' tendencies that
override caution) and exogenous (quality of the system prompt and
sandbox configuration from B2, quality of the task prompt from F1,
codebase context from F3). Timing variability can itself drive
downstream issues: very fast output creates automation bias in the
developer (F1) and reviewers (F3), reducing the perceived need for
careful scrutiny.

\paragraph{F3: To integrate changes into the codebase.}

F3 contains both automated and human sub-components with different
variability profiles. The automated sub-components (CI tests,
linters, security scanners, AI-powered PR review) have relatively
stable but bounded precision: they catch known patterns reliably but
have fixed detection ceilings for novel or subtle issues. The
AI-powered PR reviewer---an LLM reviewing another LLM's
code---introduces a potential common-mode concern, where both
systems share similar blind spots. The human sub-component (code
review) is highly variable: precision ranges from careful,
security-aware examination to cursory approval driven by time
pressure, automation bias, and anchoring by the agent's explanatory
rationale.

The principal sources of variability are endogenous (test suite
coverage, scanner detection limits, reviewer expertise and
attentiveness) and exogenous (deadline pressure, review backlog,
organisational norms around review speed). Critical upstream
couplings amplify variability at F3: if F2's output is accompanied
by a convincing rationale, the reviewer is less likely to
scrutinise closely (Pattern~4); if F4's monitor risk score has not
arrived (due to asynchronous processing), the reviewer has no
external signal to increase scrutiny; if B2's review requirements
have not been updated to reflect increased agent capability
(Pattern~2), the review process is calibrated to an outdated threat
model. F3 is the point in the system where the variability of
multiple upstream functions converges, making it the function most
susceptible to resonance effects.

\paragraph{F4: To monitor and oversee agent activity.}

F4's output---the monitor risk score---varies in both precision and
timing. Precision ranges from correctly identifying genuinely
suspicious behaviour with well-calibrated scores to missing novel
patterns (false negatives), flagging benign behaviour (false
positives that erode trust over time), or failing to cover certain
deployment surfaces entirely. The LLM-based supervisor may share
systematic blind spots with the agent it monitors, creating a
potential ceiling on detection capability that cannot be overcome by
threshold tuning alone.

Timing is a critical variability dimension for F4. Both Anthropic
and OpenAI describe monitoring as
asynchronous~\citep{openai-monitoring-2026, HttpsAlignmentanthropiccom2025},
meaning the risk score may arrive at F3 after the PR has already
been reviewed or merged. When this happens, the F4$\to$F3 coupling
provides only post-hoc detection value rather than functioning as an
input to the review decision.

The principal sources of variability are endogenous (LLM supervisor
capability limits, classifier accuracy, summarisation fidelity,
threshold calibration) and exogenous (volume of agent sessions
relative to fixed analyst budget, evolution of agent behaviour
outpacing detection rule updates, B2's threshold settings). Over
longer timescales, F4's precision is subject to the same drift
dynamics as other functions: alert thresholds are periodically
relaxed to reduce false positive burden, detection rules that have
not fired are pruned, and analyst expectations calibrate to a
baseline dominated by false positives rather than genuine threats.

\end{document}